\documentclass{aa}  

\usepackage{graphicx}

\usepackage{txfonts}
\usepackage{gensymb}

\usepackage[usenames,dvipsnames,svgnames,table]{xcolor}

\usepackage{xpatch}
\usepackage{hyperref}
\hypersetup{
	colorlinks=true,
	breaklinks=true,
	citecolor=blue,
	allcolors=blue,
	frenchlinks=true
}

\usepackage{xpatch}
\usepackage{hyperref}
\hypersetup{
	colorlinks=true,
	breaklinks=true,
	citecolor=blue,
	allcolors=blue,
	frenchlinks=true
}

\makeatletter
\xpatchcmd\NAT@citex
{%
	\@citea\NAT@hyper@{%
		\NAT@nmfmt{\NAT@nm}%
		\hyper@natlinkbreak{\NAT@aysep\NAT@spacechar}{\@citeb\@extra@b@citeb}%
		\NAT@date
	}%
}
{%
	\@citea
	\NAT@nmfmt{\NAT@nm}%
	\NAT@aysep\NAT@spacechar
	\NAT@hyper@{\NAT@date}%
}
{}{}
\xpatchcmd\NAT@citex
{%
	\@citea\NAT@hyper@{%
		\NAT@nmfmt{\NAT@nm}%
		\hyper@natlinkbreak{\NAT@spacechar\NAT@@open\if*#1*\else#1\NAT@spacechar\fi}%
		{\@citeb\@extra@b@citeb}%
		\NAT@date
	}%
}
{
	\@citea
	\NAT@nmfmt{\NAT@nm}%
	\NAT@spacechar\NAT@@open\if*#1*\else#1\NAT@spacechar\fi
	\NAT@hyper@{\NAT@date}%
}
{}{}
\makeatother

\makeatletter
\renewcommand*\aa@pageof{, page \thepage{} of \pageref*{LastPage}}
\makeatother

\begin{document} 

\newcommand{\karin}[1]{{\color{black} #1}}
\newcommand{\karintwo}[1]{{\color{black} #1}}

   \title{The diffuse gamma-ray sky of a Milky Way analog: Local diversity and global constraints}

   \author{Karin~Kjellgren
          \inst{1}
          \and
          Philipp~Girichidis\inst{1}
          \and 
          Maria Werhahn\inst{2}
          \and
          Ralf~S.~Klessen\inst{1,3}
          \and
          Christoph Pfrommer\inst{4}
          \and
          Juan Soler\inst{5}
          \and
          Brian Reville\inst{6}
          \and
          Jim Hinton\inst{6}
          \and
          Patrick Hennebelle\inst{7}
          \and
          Noé Brucy\inst{8}
          \and
          Simon C.~O.~Glover\inst{1}
          }

   \institute{Universit\"{a}t Heidelberg, Zentrum f\"{u}r Astronomie, Institut f\"{u}r Theoretische Astrophysik, Albert-Ueberle-Str.\ 2, 69120 Heidelberg, Germany
   \and
   Max-Planck-Institut f\"{u}r Astrophysik , Karl-Schwarzschild-Str. 1, 85748 Garching, Germany
   \and
   Universit\"{a}t Heidelberg, Interdisziplin\"{a}res Zentrum f\"{u}r Wissenschaftliches Rechnen, Im Neuenheimer Feld 225, 69120 Heidelberg, Germany
   \and
   Leibniz-Institut f\"{u}r Astrophysik Potsdam (AIP), 
   An der Sternwarte 16, D-14482 Potsdam, Germany
   \and
   University of Vienna, Department of Astrophysics, Türkenschanzstraße 17, 1180 Vienna , Austria
   \and
   Max-Planck-Institut für Kernphysik, Saupfercheckweg 1, Heidelberg 69117, Germany
   \and
   Université Paris-Saclay, Université Paris-Cité, CEA, CNRS, AIM,
91191 Gif-sur-Yvette, France
   \and 
   Université Lyon 1, ENS de Lyon, CNRS, CRAL, UMR 5574, Lyon, France
   }

\abstract{Diffuse gamma-ray emission is a key tracer of cosmic rays (CRs) in galaxies, encoding information about their transport, energetics, and interaction with the interstellar medium. Interpreting the Milky Way’s gamma-ray sky, however, remains challenging because the observed emission depends jointly on the \karin{three-dimensional} CR distribution \karin{and} gas distribution, \karin{as well as} the position of the observer within the Galaxy. Using the Rhea suite of CR–magnetohydrodynamic (MHD) simulations of a Milky Way analog, we investigated how pion-decay gamma-ray emission varies with galactic environment, local conditions, and CR transport physics. The emission was computed in post-processing under steady-state CR cooling and interaction assumptions, thus enabling us to analyze luminosities, spectra, full-sky emission maps, and angular power spectra (APS) for many observer positions, including those located inside Local Bubble-like cavities.
The simulated galaxy naturally reproduces Milky Way-like gamma-ray luminosities and spectral slopes without any parameter tuning. While the total luminosity remains comparatively stable across the galaxy, the detailed morphology of the gamma-ray sky varies strongly with observer location due to the complex distribution of gas in the nearby environment\karin{, consistent with longstanding observational results}. Across all observers, the APS closely follows the structure of the gas column density rather than the more diffuse CR energy density\karin{, in agreement with previous gamma-ray analyses and CR propagation models}. Comparisons with Fermi–LAT data show good agreement for both the all-sky spectrum and the APS. A diffusion coefficient energy-scaling with power-law index $\delta = 0.5$ generally matches the observations best.
\karin{Our results show that these well-established features of Galactic gamma-ray emission arise naturally in fully self-consistent CR–MHD galaxy simulations}. Gas density fluctuations are the primary drivers of the morphology of the pion-decay emission, while CR transport parameters govern its spectral and structural details. The Rhea simulations thus provide a physically grounded framework for interpreting diffuse gamma-ray observations and highlight the importance of understanding the observer’s local surroundings when using gamma rays to trace Galactic CR physics.}

   \maketitle

\section{Introduction}
Cosmic rays (CRs) are ubiquitous in galaxies; they are a crucial energy component of the interstellar medium (ISM) with energy densities that are comparable to the thermal, kinetic, and magnetic components \citep[e.g.,][]{Boulares1990,grenierNineLivesCosmic2015}. As charged particles, the CRs gyrate around and propagate in the direction of the magnetic field lines in what can be approximated as a diffusion and streaming process, allowing the particles to move relative to the thermal gas \citep{Zweibel2013, ruszkowskiCosmicRayFeedback2023}. This, combined with their inefficient cooling, makes them excellent sources for driving galactic outflows with a comparatively large mass loading factor, as seen in high-resolution setups of the ISM \citep{Girichidis2016,simpsonRoleCosmicRay2016,girichidisCoolerSmootherImpact2018,rathjenSILCCVIMultiphase2021,simpsonHowCosmicRays2023,sikeCosmicRayDrivenGalactic2024} and global galaxy models \citep{Uhlig2012,Hanasz2013,Jacob2018,buckEffectsCosmicRays2020,peschkenAngularMomentumStructure2021,monteroImpactCosmicRays2023,thomasCosmicraydrivenGalacticWinds2023,Girichidis_2024,Thomas2025}. CRs can also regulate star formation \citep{dashyanCosmicRayFeedback2020} and affect the chemistry of gas through ionization \citep[e.g.,][]{PadovaniEtAl2020}, which has already been quite extensively studied in recent numerical works \citep[e.g.,][]{HanaszEtAl2021}.

We can indirectly observe CRs via the emission that they create across a wide range of wavelengths through a variety of processes. When a CR proton interacts with thermal gas, it can create a neutral pion, which almost immediately decays into two gamma-ray photons. This is the dominant source for the diffuse gamma-ray emission in the Galaxy \citep{stecker_possible_1974,Selig_2015,Scheel-Platz_2023}, though a nonnegligible contribution (around 30\% in the 0.1--100 GeV range) comes from inverse Compton (IC) and bremsstrahlung emission from CR electrons \citep[e.g.,][]{strongGLOBALCOSMICRAYRELATEDLUMINOSITY2010,grenierNineLivesCosmic2015, Werhahn_2021_II}. Once generated, gamma rays below tens of TeV propagate freely through the ISM, unaffected by magnetic fields or absorption. As a result, the observed emission traces back to its site of origin, providing an indirect probe of the underlying CR distribution, which has motivated recent simulation-based modeling of galactic gamma-ray emission \citep[e.g.,][]{Werhahn_2021_II, werhahn2023, sandsGalacticCenterGammaRay2025}.

With instruments such as Fermi-LAT we now have observations of the all-sky diffuse gamma-ray emission in the Milky Way, as well as diffuse emission in other nearby star-forming galaxies. An observed relationship between the infrared (IR) luminosity (which probes star formation, \citealt{KennicuttEvans2012} and references therein) and gamma-ray luminosity of such galaxies has been revealed \citep{Ackermann2012,Rojas_Bravo_2016, ajello2020}, which has implications for the efficiency of CR feedback \citep{pfrommerSimulatingGammarayEmission2017,Kornecki2020,Werhahn_2021_II,werhahn2023}. Highly star-forming galaxies are found to be better proton calorimeters at a level of 60-80 percent, meaning that the CRs lose most of their energy through \karin{hadronic interactions} within the galaxy, which limits their dynamical impact.\footnote{This is in contrast to adiabatic or streaming losses, for example, where the CR losses result in thermal and dynamical impact on the gas. We note that the degree of how much energy can have a dynamical impact also strongly depends on the coupling of CRs with the gas \citep{farberImpactCosmicRayTransport2018,sikeCosmicRayDrivenGalactic2024,thomas2025_effective}.}  The Milky Way, however, is found to be only weakly calorimetric, which could be explained by efficient CR escape due to diffusion and streaming \citep{Thomas2025}. Ground-based instruments such as LHAASO, Tibet AS$\gamma$, and ARGO have also given us data of the diffuse gamma-ray spectra in different regions of the sky. Observations of our own galaxy in the last years have revealed several interesting features whose origins remain under debate, such as the GeV excess in the inner Galaxy \citep{Ackermann_2017} and the Fermi bubbles: two massive gamma-ray bubbles emanating from the Galactic Center both above and below the midplane \citep{suGIANTGAMMARAYBUBBLES2010}.

We can observe the diffuse gamma-ray emission from the Milky Way with greater resolution than any other system, and it is the only galaxy in which we can directly detect CR intensities and spectra. However, we can only observe our Galaxy from within it, and only from one single observer’s position, which complicates the derivation of global properties. Our observations are also shaped by our location inside the Local Bubble \citep{CoxReynolds1987, Zucker_2022, ONeillEtAl2024}, a cavity likely created by a series of clustered supernovae (SNe). \karin{Early gamma-ray studies based on data from the satellites SAS-2 and COS-B established the correlation between the distribution of gamma-ray emission and gas structure \citep{bignamiHighenergyGalacticGamma1975,paulDistributionGasMagnetic1976}, demonstrating that gamma-ray emission from off the Galactic plane serves as valuable tracers of local interstellar gas \citep{lebrunGammaRaysDense1978,bignamiCOSBObservationsGammaray1981}. This conclusion has also been re-established with Fermi-LAT data, based on the dominance of the gas distribution in the 8--10 kpc ring \citep{casandjianLocalHscEmissivity2015,aceroDevelopmentModelGalactic2016}. However, the role of the local environment in setting the gamma-ray sky has yet to be examined using large-scale galaxy simulations that solve the CR--magnetohydrodynamic (MHD) equations. By solving these equations, the simulations capture CR feedback on the gas and allow the resulting emission to be calculated self-consistently. This is the goal of the present work. Simulations additionally have the advantage of allowing us to study how the sky changes as observers are put in different locations within the Galactic disk.}

Our work \karintwo{is} based on the Rhea suite of hydrodynamical simulations of isolated galaxies, aimed at reproducing key characteristics of the Milky Way \citep{goeller2025}. We used a subset of the suite that includes a magnetic field and CRs \citep{kjellgren2025}. The gamma-ray emission was calculated during post-processing using the \textsc{crayon+} code \citep{Werhahn_2021_I, Werhahn_2021_II}, which calculates steady-state spectra of CR protons and electrons given the gas properties and CR energy density in every computational cell, as well as all related nonthermal emission processes; relevant for this work  is the neutral pion decay calculation from CR proton spectra. This method provides a self-consistent calculation of the emission based on the physical properties of the simulated galaxy with SNe as CR sources that derive from self-consistent star formation modeling and the detailed implementation of relevant cross sections, without fine-tuning parameters and adopting source distributions to match observations.

This paper is structured as follows. In Section \ref{sec:methods} we describe the setup of our simulations as well as our modeling of the gamma-ray emission from neural pion decay. In Section \ref{sec:gammaglobal} we  present our results by showing the morphology of the gamma-ray emission, and how well we reproduce global estimates of the luminosity of the Milky Way. Section \ref{sec:localemission} investigates the gamma-ray sky as seen by different observers, in particular the role of the local environment on the emission. We investigate the correlation between the gamma-ray emission and the gas and CR energy distribution in Section \ref{sec:gasandcrenergy}, and in Section \ref{sec:observations} we compare the simulated gamma-ray sky with observations of the Milky Way. We present our discussion in Section \ref{sec:disc} and conclude in Section \ref{sec:conclusions}.

\section{Methods}\label{sec:methods}
\subsection{Simulation setup}\label{subsec:simsetup}
We simulated isolated galaxies using the moving-mesh code \textsc{arepo} \citep{springelPurSiMuove2010,Pakmor2016,arepo_public}. Full details of the setup are given in \cite{kjellgren2025} and \cite{goeller2025}; here we summarize the most important points.

We started with a smooth gaseous disk with a gas mass of $1.2\times10^{10}~\mathrm{M_\odot}$. To represent the gravitational influence of stars, we used an external flat gravitational potential that results in a flat velocity curve. The gas mass resolution is 3000 $\mathrm{M}_\sun$, and the minimum and maximum allowed cell volumes are 1 pc$^3$ and 2 kpc$^3$, respectively. In the disk ($r_{xy}\leq30$ kpc, $h\leq1$ kpc) there is additional volume refinement that limits the maximum cell volume to (100 pc)$^3$. Stars are formed using collisionless star particles, each of which represent a stellar population. Star particles are created based on the gas mass of the cell, either deterministically based on its Jeans mass, or stochastically as in the \cite{springelCosmologicalSmoothedParticle2003} ISM model, and are populated based on a Kroupa IMF using the algorithm in \cite{sormaniSimpleMethodConvert2017}. High-mass stars ($8~\mathrm{M_\odot}<M_\star<120~\mathrm{M_\odot}$) eventually explode as SNe, injecting momentum and energy into \karin{the surrounding 100 pc}. We let the simulations evolve for 2 Gyr. During the entire evolution, mass return from SNe is activated, meaning that after the SNe explode, all their mass is returned to the neighboring cells. This has the consequence that no mass is locked up in the star particles, and that we do not deplete the gas. The simulations presented in \cite{goeller2025} disable mass-return after the first two gigayears of evolution;  however, due to the computational cost of the CR runs we only focused on this phase for this work.

\karin{The Rhea simulations use the NL97 nonequilibrium chemical network from \cite{gloverApproximationsModellingCO2012}, which tracks the abundances of atomic, ionized, and molecular hydrogen, as well as a simplified treatment of carbon and CO. These are used as inputs for the atomic and molecular heating and cooling function described in \cite{clarkTracingFormationMolecular2019}, where a detailed description of the included processes can be found in \cite{gloverModellingCOFormation2010}, with later modifications made by \cite{gloverApproximationsModellingCO2012} and \cite{mackeyNonequilibriumChemistryDestruction2019}}.

The magnetic field is initialized as a purely toroidal field, with an initial magnetic field strength of either 3 $\mu$G (CRMHD) or 3 nG (CRMHD-low). A detailed comparison of these two simulations can be found in \citet{kjellgren2025}, but for the analysis in this paper we purely focus on the CRMHD simulation. To show the robustness of our results we show some of the analysis in this paper applied to CRMHD-low in Appendix \ref{app:crmhdlow}.

Cosmic rays are included in the gray approximation, where we only evolve the total integrated CR-proton energy density \citep{PakmorEtAl2016, pfrommerSimulatingCosmicRay2017}. CR energy is injected \karin{for each SN} as 10\% of the explosion energy, \karin{and is deposited into the same cells as the thermal energy. They are} transported in the advection-diffusion approximation with a constant diffusion coefficient of $D_{0} = 4\times10^{28}\;\mathrm{cm}^2\mathrm{s}^{-1}$ directed along the magnetic field. The following losses of CR energy are accounted for in \textsc{arepo}: hadronic losses, Coulomb losses, adiabatic losses, and Alfvén losses (which emulate the losses due to streaming; see \citealt{Wiener2017}).

\subsection{Modeling emission with \textsc{crayon+}}\label{subsec:crayon}
Gamma-ray emission from pion decay is calculated in post-processing using \textsc{crayon+}, a code that calculates steady-state CR and gamma-ray spectra in every cell of the simulation. Details of how \textsc{crayon+} works can be found in \cite{Werhahn_2021_I,Werhahn_2021_II}; here we summarize the most important points and assumptions.

\subsubsection{Steady-state CRp spectra}
The steady-state spectrum $f(E)=\mathrm{d}^2N/(\mathrm{d}E~\mathrm{d}V)$ is calculated for CR protons by solving the diffusion-loss equation, which means solving
\begin{equation}
    \frac{f(E)}{\tau_{\rm esc}}-\frac{\mathrm{d}}{\mathrm{d}E}[f(E)b(E)]=q(E),
\end{equation}
where $E$ is the CR proton energy, and $b(E)$ and $q(E)$ are losses and sources, respectively. The escape of CRs due to diffusion and advection is included in the escape timescale $\tau_{\rm esc}^{-1}=\tau_{\rm diff}^{-1}+\tau_{\mathrm{adv}}^{-1}$, where $\tau_{\mathrm{diff}}$ and $\tau_{\mathrm{adv}}$ are defined using the CR gradient length $L_{\rm CR}=e_{\rm CR}/|\boldsymbol{\nabla}e_{\rm CR}|$, where $e_{\rm CR}$ denotes the CR energy density, as
\begin{equation}\label{eq:t_esc}
    \tau_{\rm diff}=\frac{L_{\rm CR}^2}{D(E)},~~~~\tau_{\rm adv}=\frac{L_{\rm CR}}{\varv_z}.
\end{equation}
\karin{For the advection timescale only the velocity in the $z$~direction, $\varv_z$, is used because azimuthal velocities entering and leaving each cell are assumed to cancel each other out \citep[see figure~6 of][and surrounding discussion]{Werhahn_2021_I}.} 

Higher-energy CRs  diffuse faster \citep[e.g.,][]{evoliAMS02BerylliumData2020}, thereby modifying the resulting CR spectrum as well as the subsequent gamma-ray spectrum. Although we only evolve the total integrated CR energy in the simulation, and use a constant value for the diffusion coefficient, \textsc{crayon+} is able to account for an energy-dependent diffusion coefficient $D(E)=D_0(E/E_0)^\delta$, where $E_0=3~\rm GeV$ and $D_0$ is the diffusion coefficient used in \textsc{arepo}. We chose a scaling of the CR diffusion coefficient of $\delta=0.5$ as our fiducial value, though we also tested a shallower scaling of $\delta=0.3$. Changing $\delta$ should in theory also affect the dynamics in the simulation, but to account for that we would need to evolve the full CR spectra in an energy-dependent way, which we do not do here, but leave for future work. For the energy-dependent losses $b(E)$ hadronic and Coulomb losses are accounted for, and the source spectrum $q(E)\propto p(E)^{-\alpha_{\rm inj}} {\rm d}p/{\rm d} E$ is assumed to be a power law in momentum with a spectral index of $\alpha_{\rm inj}=2.2$ \citep{lackiDIFFUSEHARDXRAY2013}. After the effects of losses and escape have been applied, the resulting CR spectra are re-normalized based on the CR energy density of each cell. The Alfvén losses are implicitly included when we re-normalize the spectra but are not included as an energy-dependent loss, even though this should also be a function of CR energy. However, a detailed modeling of spectrally resolved CRs and a subsequent analysis of the relative energies reveals that the Alfv\'{e}n losses at above $\sim100\,\mathrm{GeV}$ do not contribute much to the overall energy budget \citep{Girichidis_2024}.

The code assumes steady-state, meaning that CR losses and sources balance. \cite{Werhahn_2021_I} investigated the validity of this assumption and found that the cells that predominantly contribute to the gamma-ray emission obey the conditions for steady-state. The assumption breaks down in some regions, however, in particular in the outflow region and close to the CR injection sites. Moreover, \citet{werhahn2023} compared the gamma-ray emission characteristics derived from a spectral CR simulation of an isolated galaxy \citep{Girichidis_2020,Girichidis_2022,Girichidis_2024} to steady-state gamma-ray spectra derived with \textsc{crayon+} and found excellent agreement after adjusting the CR diffusion coefficient. \karin{We also validate this assumption in the context of the Rhea simulations in Appendix \ref{app:steadystate}.}

\begin{figure*}
    \centering
    \includegraphics[width=\textwidth]{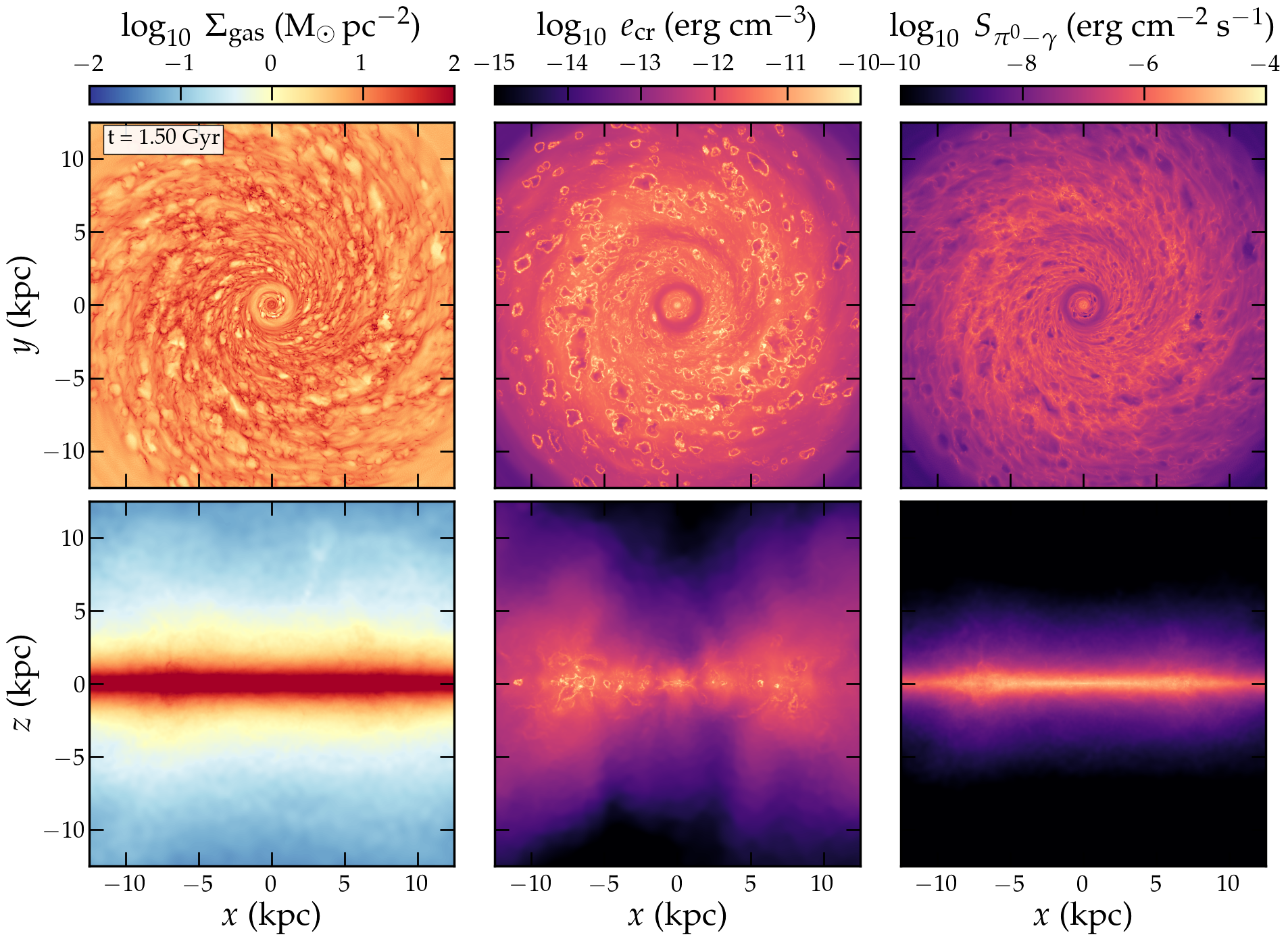}
    \caption{Face-on and edge-on views of our simulation CRMHD at $t=1.50$ Gyr. From left to right: Column density, a slice through the midplane of the CR energy density, and the projected gamma-ray emissivity from pion decay. Almost all of the mass is located in the galactic midplane. CR-driven outflows push gas out into the CGM, but of significantly lower column densities than the disk. The CRs diffuse out of the galactic disk and fill the CGM, where they establish pressure gradients capable of launching outflows. The resulting pion-decay gamma-ray emissivity is confined close to the midplane, since the low density in the CGM reduces the amount of possible gas targets for the CRs.}
    \label{fig:maps}
\end{figure*}

\subsubsection{Gamma-ray emission}
Inelastic collisions between CR protons and other gas protons produce neutral pions, which decay into gamma-ray emission. The source function $q_\gamma(E)=\mathrm{d}^3N_\gamma/(\mathrm{d}V~\mathrm{d}t~\mathrm{d}E)$ of the gamma-ray emission from neutral pion decay is calculated in \textsc{crayon+} using the CR proton distribution described in the section above, and the parameterization of the cross section by \cite{yangExploringShapeRay2018} from the pion production threshold ($p>0.78~\mathrm{GeV}/c$) up to 10 GeV, and by \cite{kafexhiuParametrizationGammarayProduction2014} for larger proton kinetic energies. This yields the gamma-ray emissivity $j_{E,\pi^0}=Eq_\gamma$, in every cell. Once again we refer to \cite{Werhahn_2021_II} for details.

We only include gamma-ray emission from the decay of neutral pions generated in collisions between CR protons and gas particles, emission in the form of IC and bremsstrahlung from primary and secondary electrons is not included. Since electrons undergo  more rapid losses than the protons, it is crucial to properly model the electron population in a non-steady-state way. In particular, IC emission in the energy range 0.1--100 GeV originates from electrons with different Lorentz factors, depending on the incident radiation field. For example, for IR photons (of energy ${\sim}10^{-2}$ eV), the required electron normalized momenta for IC scattering into the gamma-ray regime (0.1--100 GeV) are $\gtrsim 10^5$, which are affected by rapid losses and deviate from a steady-state \citep{Werhahn2025}. A proper implementation of the IC emission would also require careful modeling of the radiation field. Nevertheless, at the energies considered in this work, pion decay likely dominates the gamma-ray budget \citep[e.g.,][]{Lacki2011,Werhahn_2021_II}. For these reasons we chose to only focus on the gamma-ray emission from pion decay and leave the other components for future work. We discuss the potential impact of the missing electrons in Section~\ref{sec:disc}. For the generation of Mollweide projections of the gamma-ray sky in our simulation, as well as calculation of the APS, we  made use of the \texttt{healpy} and \texttt{HEALpix} packages \citep{gorskiHEALPixFrameworkHighResolution2005,zoncaHealpyEqualArea2019}.

\begin{figure}
    \centering
    \includegraphics[width=\columnwidth]{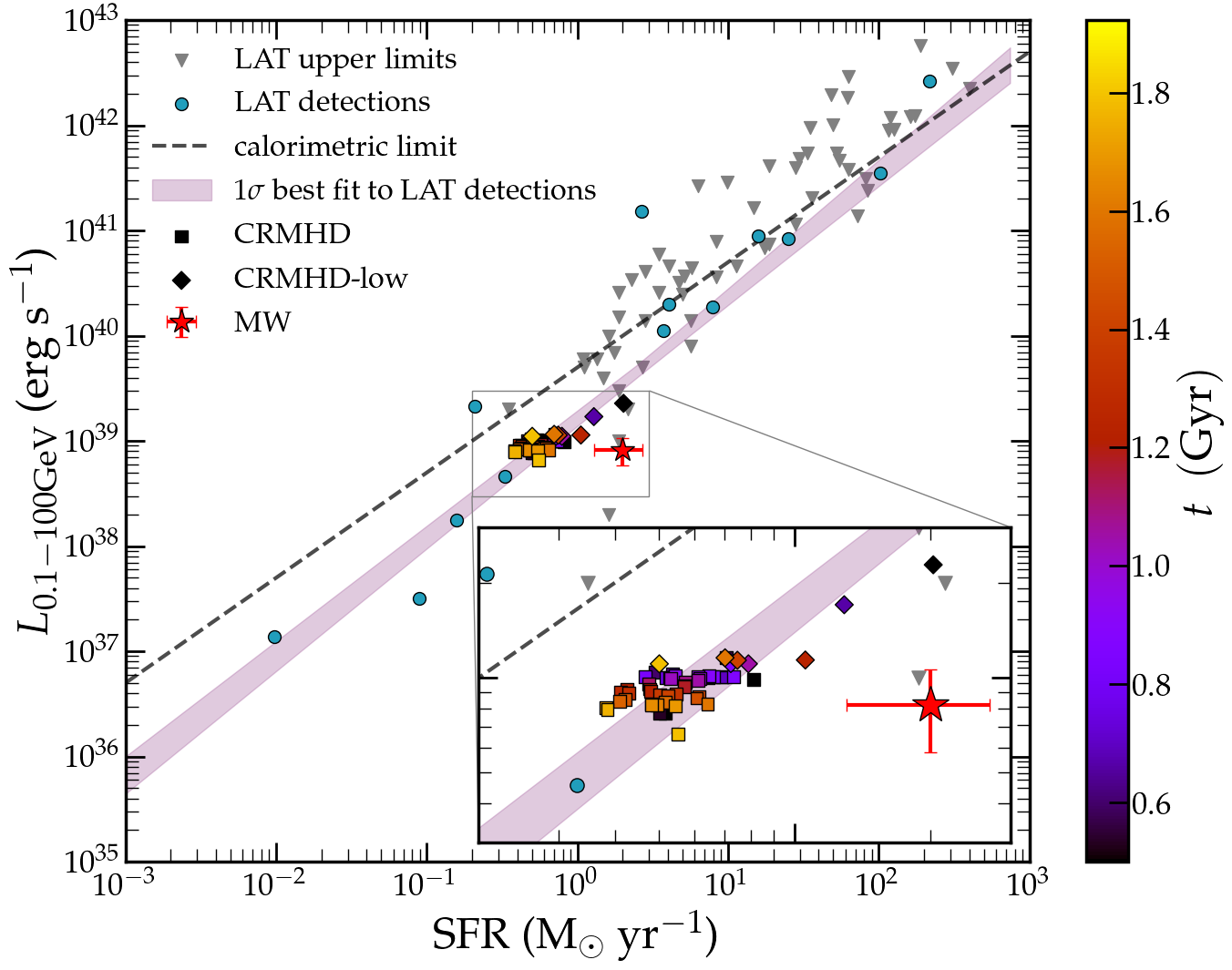}
    \caption{Relation between SFR and gamma-ray luminosity in the 0.1-100~GeV energy band for our simulations. The light blue circles are LAT detections from \cite{ajello2020} and the gray triangles are LAT upper limits from \cite{Rojas_Bravo_2016}. The Milky Way gamma-ray luminosity is from \cite{Ackermann2012}, while the SFR is  from \cite{eliaStarFormationRate2022}. The shaded purple band shows the 1$\sigma$ best fit to the LAT detections \citep{ajello2020}. The square and diamond markers show the simulations CRMHD and CRMHD-low respectively, color-coded by the time of the snapshot. The zoomed-in region in the inset more clearly shows how the simulations change with time.}
    \label{fig:fcal}
\end{figure}

\begin{figure}
    \centering
    \includegraphics[width=\columnwidth]{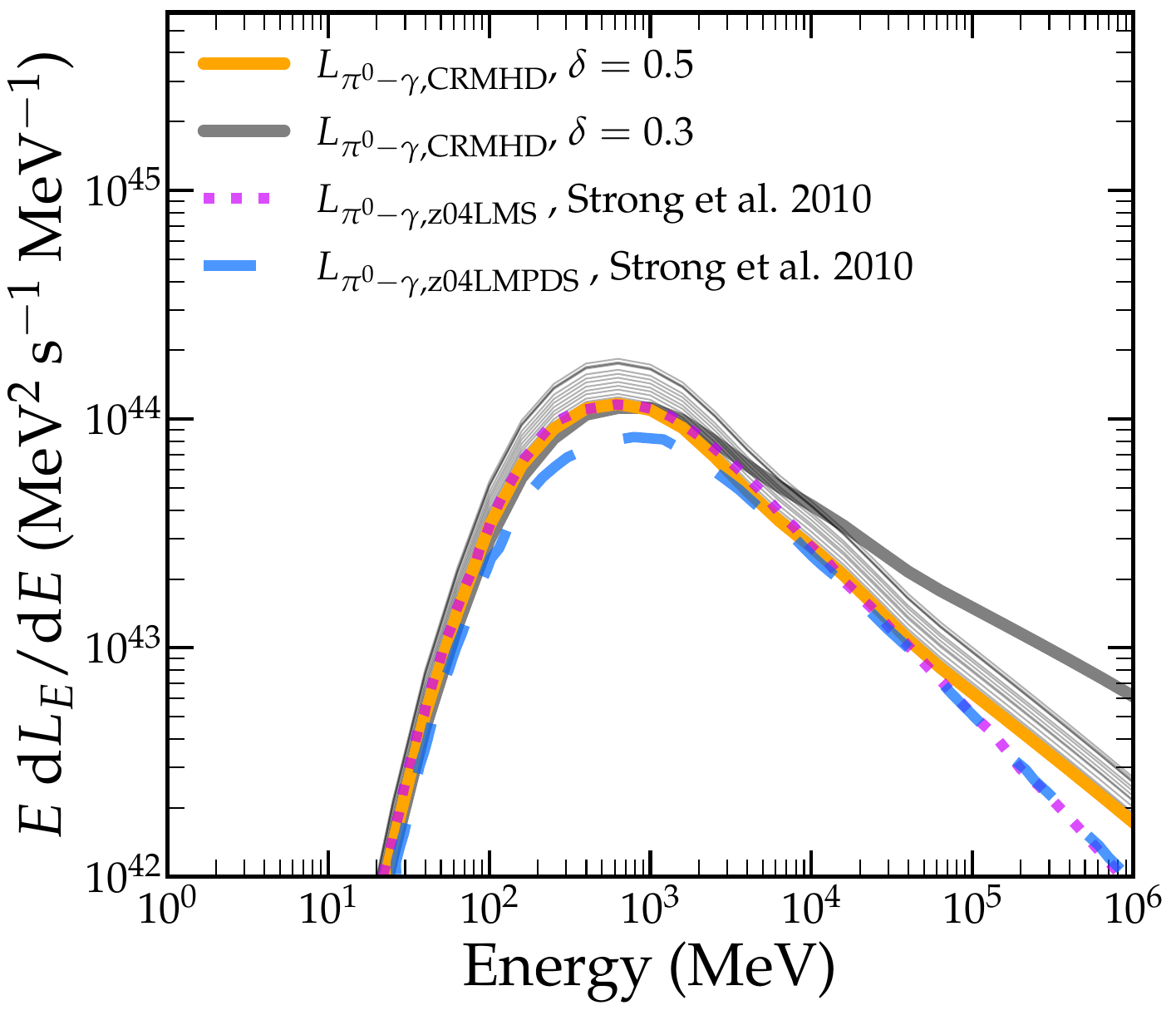}
    \caption{Estimates of the luminosity spectrum from neutral pion decay of the Milky Way from two different propagation models (dotted and dashed lines) \citep[see Fig.~1 in ][]{strongGLOBALCOSMICRAYRELATEDLUMINOSITY2010}. The solid colored lines show our simulated gamma-ray spectrum from our closest matching snapshot ($t=1.92$ Gyr). Orange depicts our fiducial \textsc{crayon+} run with $\delta=0.5$; the thick gray line shows $\delta=0.3$ instead. The thin gray lines show the luminosity spectrum for every fifth snapshot ($\Delta t\approx25$ Myr) between $t=1.5$ Gyr and $t=2.0$ Gyr using $\delta=0.5$.}
    \label{fig:lum_spec}
\end{figure}

\begin{figure}
    \centering
    \includegraphics[width=0.95\columnwidth]{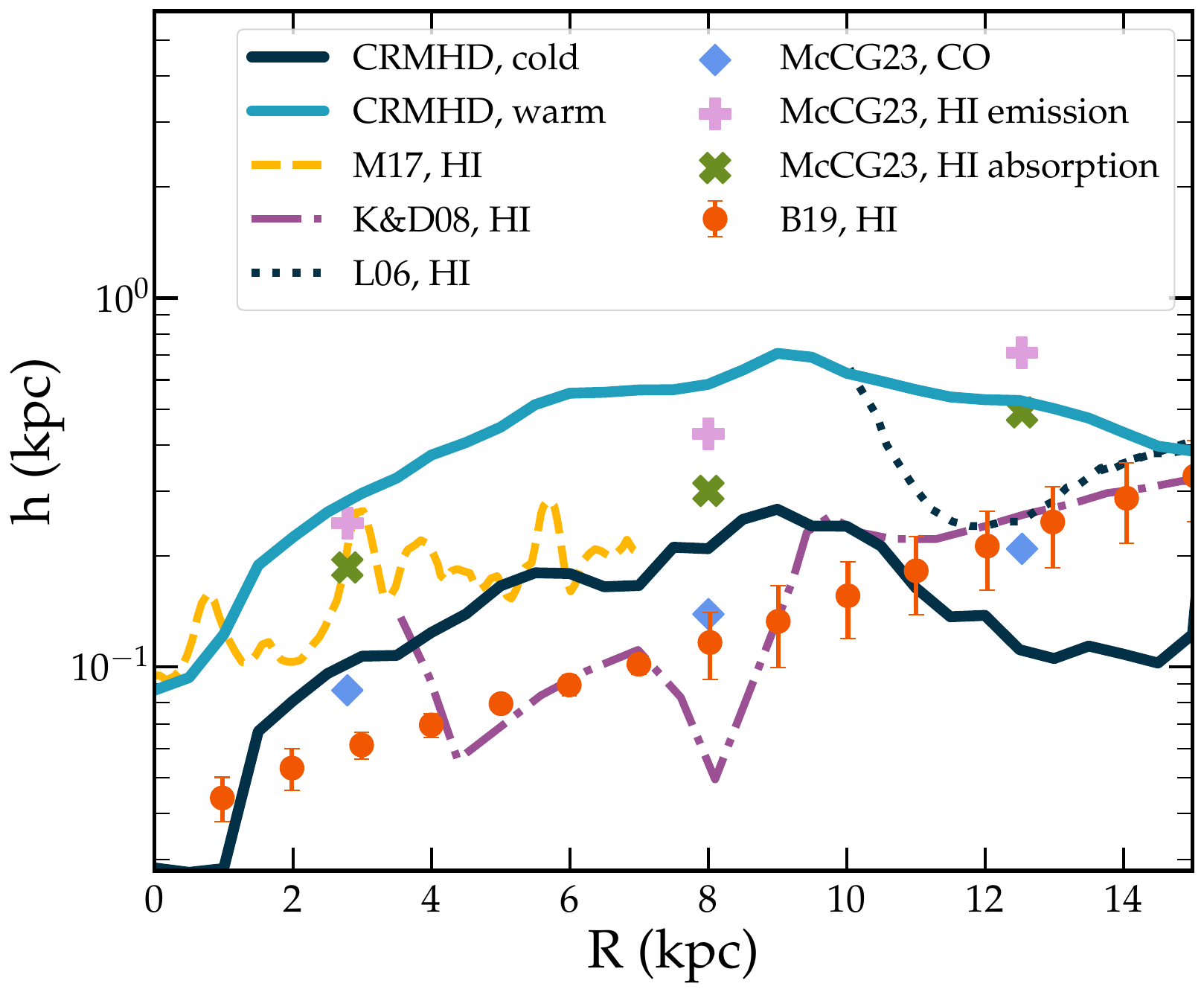}
    \caption{\karin{Comparison of the vertical scale height in our simulation compared to Milky Way observations. The solid lines show the scale height of the simulation at $t=1.92$~Gyr, calculated as 75\% of the gas mass within a certain radial bin, in the cold ($T<5050\;\rm K$) and warm ($5050\;\mathrm{K}<T<2\times10^4\;\mathrm{K}$) phase. All other data points are based on observations of the Milky Way: L06 = \cite{levine_vertical_2006}, K\&D08 = \cite{kalberla_global_2008}, M17 = \cite{marascoDistributionKinematicsAtomic2017}, B19 = \cite{bacchini_volumetric_2019}, and McCG23 = \cite{mcclure-griffiths_atomic_2023}.}}
    \label{fig:scaleheight}
\end{figure}

\begin{figure*}
    \centering
    \includegraphics[width=\textwidth]{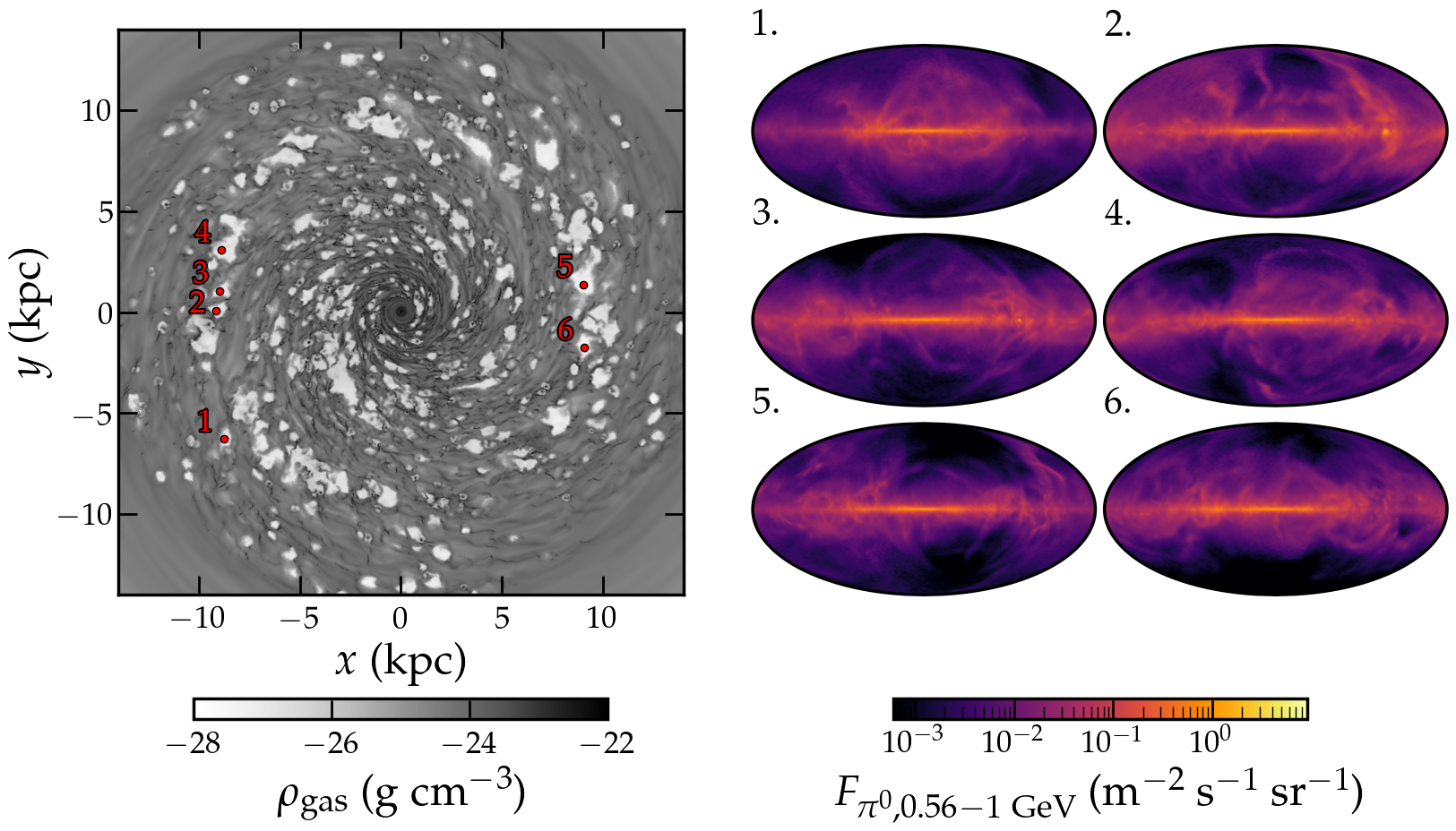}
    \caption{All-sky gamma-ray emission for different observers at the same moment in time ($t=1.92$ Gyr). The left panel shows a slice through the midplane of the gas density with the positions of the observers marked in red. The right panels show Mollweide projections of the gamma-ray emission from pion decay in the energy range 0.56--1.0 GeV.}
    \label{fig:bubbles}
\end{figure*}

\section{Gamma-ray emission in the Rhea simulations}\label{sec:gammaglobal}

Figure~\ref{fig:maps} shows face-on and edge-on projections of the gas density and the gamma-ray emissivity from pion decay, and a slice through the midplane of the CR energy density at $t=1.50~\mathrm{Gyr}$. The CR-driven outflows transport gas from the disk into the circumgalactic medium (CGM), which is also as a result infused with CR energy. The pion-decay gamma-ray emission is more confined to the galactic disk, being limited by the low gas densities in the outflow region. The face-on view reveals many small-scale details and local variations in the emission. Explosions from SNe drive expanding low-density bubbles, which become outlined by an increased amount of CR energy density. These features are also imprinted in the gamma-ray emission.

To reassure ourselves that the post-processed emission is reasonable, we would like to compare our simulations with estimations of the Milky Way gamma-ray luminosity from the literature. This is done in Fig. \ref{fig:fcal}, which depicts the well-known relationship between the gamma-ray luminosity and the star formation rate (SFR) (probed by the IR emission) for a selection of star-forming galaxies detected with Fermi-LAT \citep{ajello2020}, as well as some Fermi upper limits \citep{Rojas_Bravo_2016} that could potentially house AGNs. The luminosities (gamma-ray, IR) of the Milky Way, whose properties are what we primarily would like to reproduce, were taken from \cite{Ackermann2012} and are based on a numerical model of CR transport and ISM interaction, operated on static Galactic models \citep{strongGLOBALCOSMICRAYRELATEDLUMINOSITY2010}. The SFR of the Milky Way is often quoted as between $1-2~\rm M_\odot\ yr^{-1}$ \citep{2011Chomiuk,Licquia_2015,eliaStarFormationRate2022}; here we   used the value and uncertainty obtained in \cite{eliaStarFormationRate2022} based on observational Herschel data from the entire Galactic disk. There are, however, estimates as low as $0.67~\rm M_\odot \ yr^{-1}$ based on observations of local high-mass stars \citep{2025Quintana}, which is closely in line with our simulations, but could be underestimating the amount of star formation in the Galaxy by extrapolating from the local ISM. Nonetheless, we acknowledge that the exact value of the SFR of the Milky Way is uncertain. 

Our gamma-ray luminosities are consistent with the Milky Way. This particular energy range (0.1-100 GeV) is likely dominated by the pion-decay component \citep{Selig_2015,Scheel-Platz_2023} and should at most have a small contribution from CR electrons, which we do not include. We note that the temporal variations in the simulations are small, indicating that our galaxy is in dynamical equilibrium. The SFR in our simulations is on the lower end compared to the Milky Way, resulting in a degree of calorimetry (i.e., distance to the calorimetric limit, dashed line) higher than expected for the Galaxy. This could suggest that our CR energy injection efficiency of 10\% is too high, i.e., that we produce too many CRs considering our SFR. \cite{paisEffectCosmicrayAcceleration2018}, for example, found that an average efficiency of 5\% is more reasonable for SNe expanding in a turbulent magnetic field with varying orientation with respect to the blast wave. However, considering the aforementioned spread in observational values of the SFR, our data remain consistent with current estimates.

In Fig.~\ref{fig:lum_spec} we further compare the luminosity spectrum of the full galactic disk from the CRMHD simulation with two different estimates for the Milky Way from \citet{strongGLOBALCOSMICRAYRELATEDLUMINOSITY2010}, which provides one of the most widely accepted models of the Galaxy's gamma-ray luminosity \citep[e.g.,][]{grenierNineLivesCosmic2015}. The authors used the GALPROP code to calculate the luminosity spectrum from neutral pion decay (along with the other components) based on different CR propagation models, calibrated to match the direct observations of Fermi-LAT. The two models predict slightly different luminosities since the underlying transport parameters differ \karin{(the LMS model uses $\delta=0.33$ and the LMPDS model uses $\delta=0.5$)}, but the spectral shapes are similar, in particular the slopes. The thin gray lines show the luminosity spectra for every fifth simulation snapshot ($\Delta t\approx25$ Myr) between $t=1.5$ Gyr and $t=2.0$ Gyr. There are minor variations in the total luminosity with time, as was evident from Fig. \ref{fig:fcal}, but the spectral shape remains constant. The solid orange and gray lines correspond to a snapshot at $t=1.92$ Gyr using either a scaling of $\delta=0.5$ or $\delta=0.3$ in \textsc{crayon+}, respectively. The amplitude of the spectrum at this time matches the z04LMS-model of \cite{strongGLOBALCOSMICRAYRELATEDLUMINOSITY2010} particularly well (see Fig.~\ref{fig:lum_spec}). Regardless of the GALPROP model, adopting a scaling of $\delta=0.3$ in \textsc{crayon+} clearly overestimates the luminosity at higher energies ($\gtrsim1~\rm GeV$). In contrast, the model with $\delta=0.5$ reproduces the high-energy slope more accurately, with only a small deviation at $E\gtrsim10^5$ MeV. We conclude that although it somewhat depends on the model of CR propagation, we manage to reproduce the luminosity spectrum of the Milky Way well for the snapshot at $t=1.92~\mathrm{Gyr}$ with $\delta=0.5$, which we  use as our fiducial values. 

We note that the more efficient production of gamma-ray emission in our simulations compared to the Milky Way, evident in Fig.~\ref{fig:fcal}, may partly reflect the differences between our setup and the GALPROP models of \citet{strongGLOBALCOSMICRAYRELATEDLUMINOSITY2010}. In particular, those models assume purely isotropic diffusion, enabling more a efficient escape of CRs than in Rhea, which adopts parallel diffusion in a predominantly toroidal magnetic field. Additional discrepancies in assumptions regarding CR injection and the gas density distribution further complicate a direct comparison, which we therefore defer to future work.

\karin{Before proceeding with the analysis, we also compare the vertical scale height in our simulated galaxy with observational estimates of the Milky Way in Fig. \ref{fig:scaleheight} as this is important for the resulting gamma-ray sky and the role of the local environment. The scale height is calculated in axisymmetric radial bins as the height that contains 75\% of the gas mass, and is separated into the cold ($T<5050$~K) and warm ($5050\;\mathrm{K}<T<2\times10^4\;\mathrm{K}$) phase \citep{kimNumericalSimulationsMultiphase2018}. Estimations of the Milky Way gas scale height from the literature are shown for comparison, primarily based on HI data, and using a variety of different methods. Though we do not see the same amount of flaring at large radii as suggested by HI observations \citep[e.g.,][]{kalberla_global_2008}, the scale height in our simulation is well in line with observational estimates.}

\begin{figure}
    \centering
    \includegraphics[width=\columnwidth]{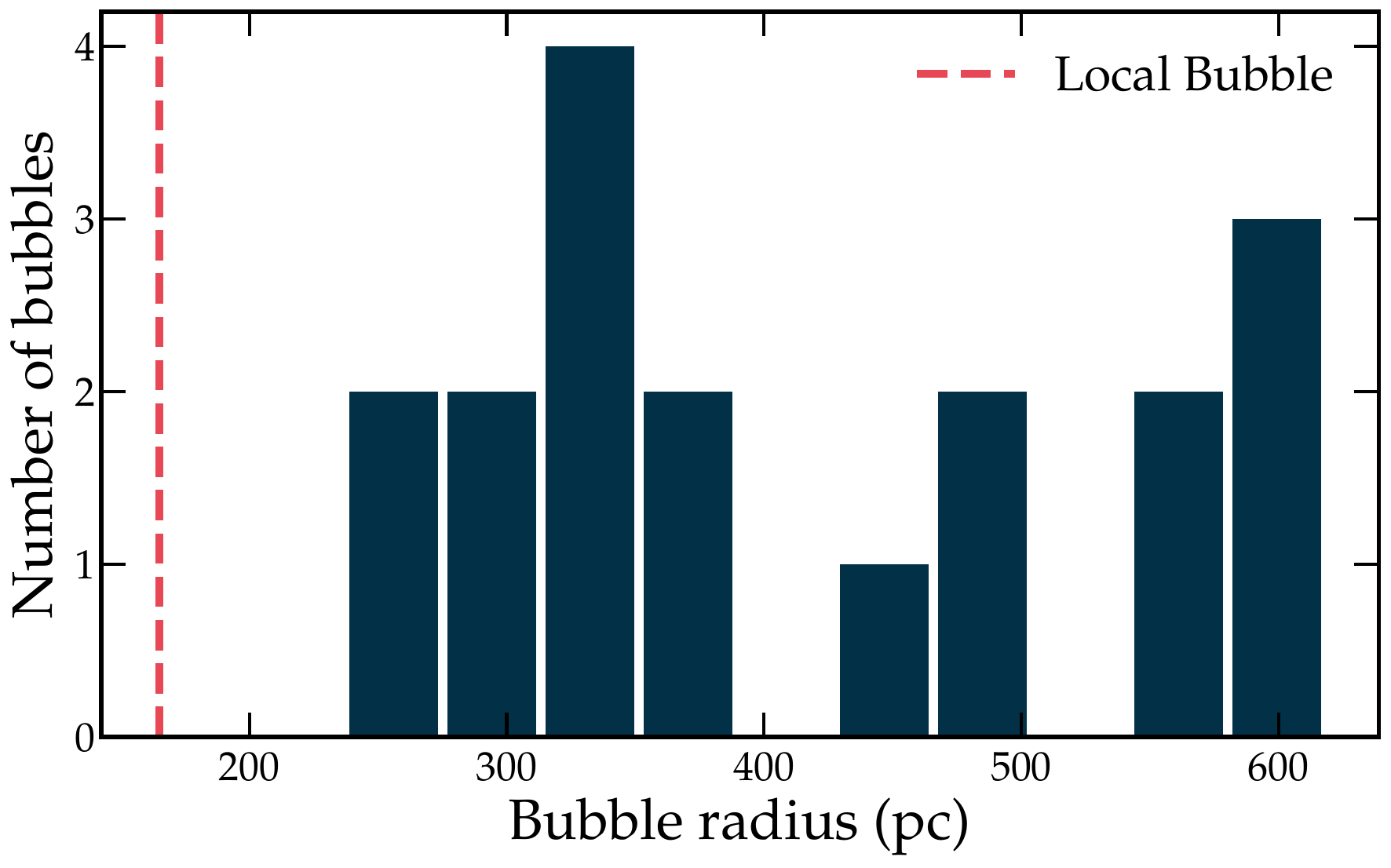}
    \caption{\karin{Histogram of the circularized bubble radii of 18 bubbles selected from one snapshot in our simulation, all from within 2.5 kpc of the solar circle. The dashed line shows the radius of the Local Bubble.}}
    \label{fig:r_hist}
\end{figure}

\begin{figure*}[htbp]
  \centering
  \includegraphics[width=\textwidth]{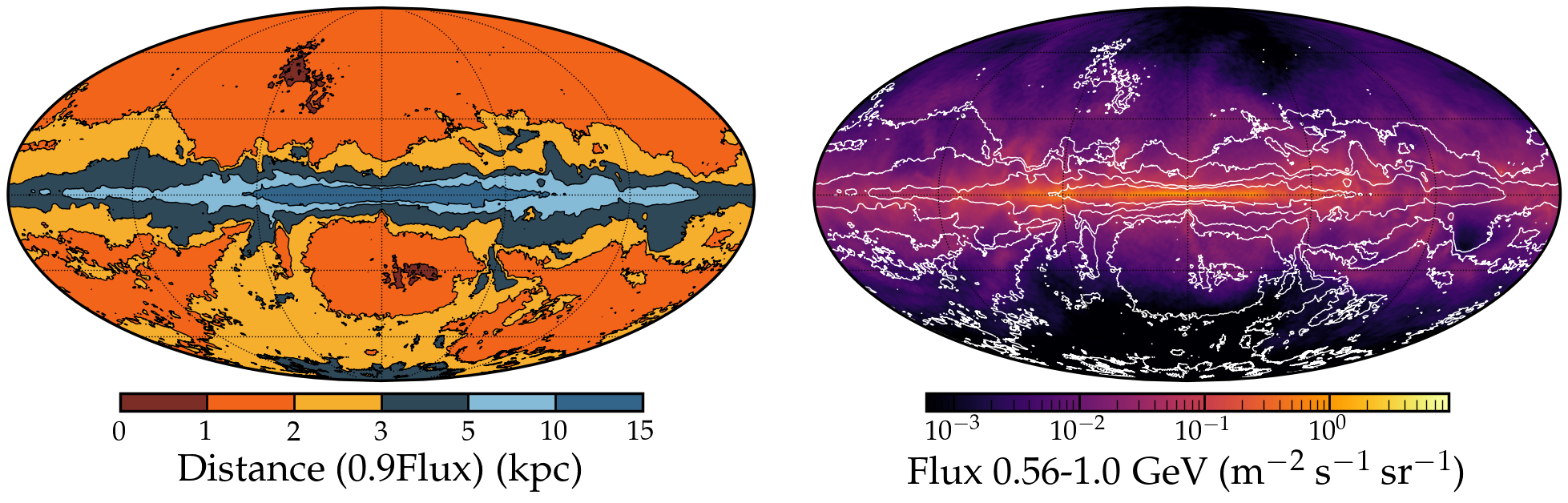}
  \caption{Left panel: For bubble 6 in Fig. \ref{fig:bubbles}, distance from which up to 90\% of the emission in each line of sight originates.
 Right panel: Corresponding gamma-ray flux with the contours of the left panel overlaid. At higher galactic latitudes the gamma-ray sky is dominated by local ($\leq2~\mathrm{kpc}$) emission.}
  \label{fig:d90_flux}
\end{figure*}

\begin{figure}
    \centering
    \includegraphics[width=\columnwidth]{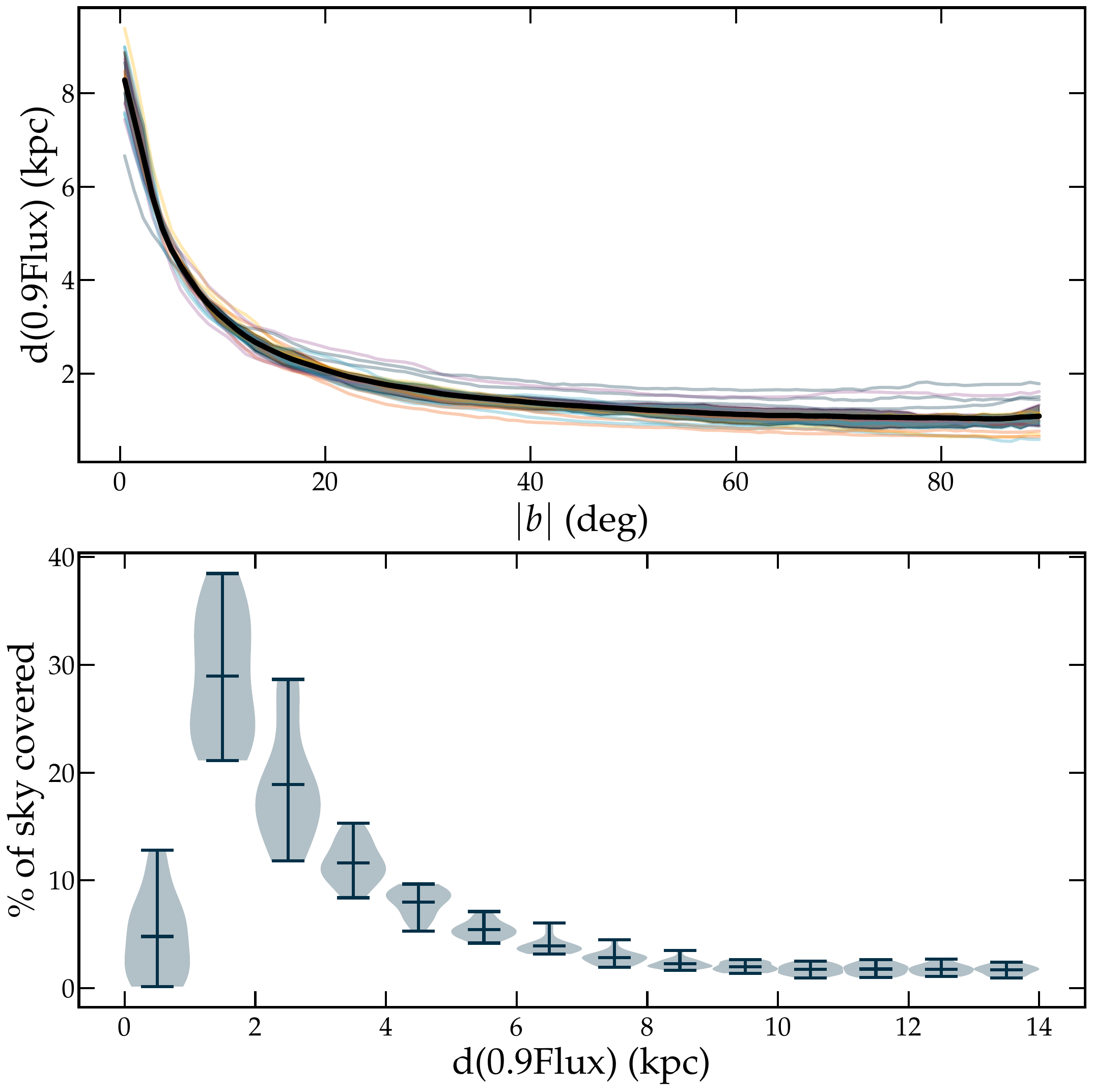}
    \caption{Top: Distance within which 90\% of the gamma-ray emission originates, as a function of galactic latitude, for a selection of 18 bubbles in one snapshot of our simulation. The black solid line shows the median evolution, and the shaded region marks the 25th and 75th percentiles. Bottom: Sky covering fraction of regions dominated by emission from different distances, in 1 kpc bins.}
    \label{fig:d90_b}
\end{figure}

\section{Variations in the gamma-ray sky and the role of local emission}\label{sec:localemission}

The left panel of Fig. \ref{fig:bubbles} depicts a density slice through the midplane of our simulation at $t=1.92$ Gyr, the time which provided the best match to the luminosity spectrum of the Milky Way in Fig. \ref{fig:lum_spec}; the numbered dots represent the different locations of observers. The locations were all selected to be in low-density bubbles, mimicking the fact that the Sun is inside the Local Bubble \citep{Zucker_2022}. The six panels on the right show what the gamma-ray sky (from pion decay of CR protons, in the energy range 0.56-1.0 GeV) looks like for different observers at these locations. Once gamma rays are created they do not interact, meaning it is straightforward to sum up the contributions of the individual cells to create all-sky maps. The sky projections have been rotated so they all look at the galactic center head-on, and the dynamic range and unit of the flux have been chosen to match the reconstructed diffuse dust-correlated gamma-ray sky based on ten-year Fermi data, presented in Fig. 14 of \citet{Scheel-Platz_2023}.

The bubbles in the simulation were chosen based on visual inspection and not in an attempt to find the one bubble that most closely resembles the Local Bubble. The Local Bubble that we ourselves reside in is estimated to have a radius of around 165 pc, however with large variations in different directions, in particular toward the top and bottom \citep{ONeillEtAl2024}. \karin{In Fig.~\ref{fig:r_hist} we show a histogram of the circularized radii of the 18 bubbles selected for the analysis in this work; all have galactocentric radii between 6.5 kpc and 10.8 kpc, so as to be comparable to the location of the Sun at 8.25 kpc. The radii of the bubbles from the simulation are all $2-4$ times larger than the average radius of the Local Bubble. Although our bubbles are larger, we do not believe this compromises our analysis.} For some comparisons of the gamma-ray emission in bubbles versus randomly selected positions in the disk, we refer to Appendix~\ref{sec:app1}. 

Visually, there are striking variations between the projections. Filaments and loops dominate the sky away from the galactic midplane, resulting in widely different skies even between bubbles 2 and 3, which are relatively close to each other. This suggests that the local environment plays an important role in determining the distribution of the out-of-plane emission. To center our discussion, we focus on a single bubble for much of the remaining analysis presented in this paper; specifically, we chose bubble 6 from Fig.~\ref{fig:bubbles}. We refer to Appendix~\ref{sec:app1} for some additional comparisons between the bubbles.

In order to further understand the importance of the local environment, it is useful to quantify the range of distances from which the majority of the observed emission originates. Figure~\ref{fig:d90_flux} shows, in the left panel, the maximum distance that one needs to travel along each line of sight in order to account for 90\% of the observed emission along that line of sight. For example, the orange regions in this plot denote areas of the sky for which this maximum distance is only 2~kpc. The right panel shows the flux map with the contours of the distance map overlaid. Most of the emission naturally comes from the galactic midplane, and we note that we need to include emission from distances of up to $d(0.9~\mathrm{Flux}) = 5-10$ kpc from the observer to capture 90\% of the flux in these regions, and as far as $15~\mathrm{kpc}$ in the center. With increasing galactic latitude there is less overall flux, and the majority of the emission originates from gas that is much closer, in many cases less than 2~kpc away.
A comparison of the left and right panels shows that there is clearly visible gamma-ray emission in the flux map that overlaps with the 1--2 kpc distance-bin, implying that the emission is a result of the local environment where the observer is found. 

Figure~\ref{fig:d90_b} shows how $d$(0.9 Flux) varies with galactic latitude for a larger sample of 18 bubbles from the same snapshot, again selected based on visual inspection. We see, for example, that at a galactic longitude of $20\degree$, 90\% of the emission comes from within 2 kpc of the observer. There is not a large overall variation between different the observers; statistically, the bubbles are similar and show that with increasing latitude, local emission becomes more important. \karin{This trend is a direct imprint of the vertical scale height of the gas, as portrayed in Fig. \ref{fig:scaleheight}, which limits the path length through dense material for lines of sight away from the midplane. Because the gamma-ray emission from pion decay traces dense structures, it is likewise confined close to the midplane (see Fig.~\ref{fig:maps}, bottom right panel)}. The picture could conceivably change if we were to include the leptonic IC emission, which \karin{would affect the sky} at high galactic latitudes and higher gamma-ray energies as it traces the hot outflowing gas \citep{Selig_2015,Scheel-Platz_2023}. However, at these energies, pion decay gamma rays still dominate. The bottom panel shows how much of the sky is dominated by emission from specific distance ranges. For example, over the 18 selected bubbles, the median sky fraction   dominated by emission from 2--3 kpc away is $\sim19\%$. Emission from $1\!-\!2~\mathrm{kpc}$ away dominates the sky more than any other distance bin, though the exact fractions vary between bubbles. In some cases, a nonnegligible portion of the sky is dominated by emission arising from within the nearest kiloparsec. 

\section{Correlation with gas density and CR energy density}\label{sec:gasandcrenergy}

\begin{figure}
    \centering
    \includegraphics[width=\columnwidth]{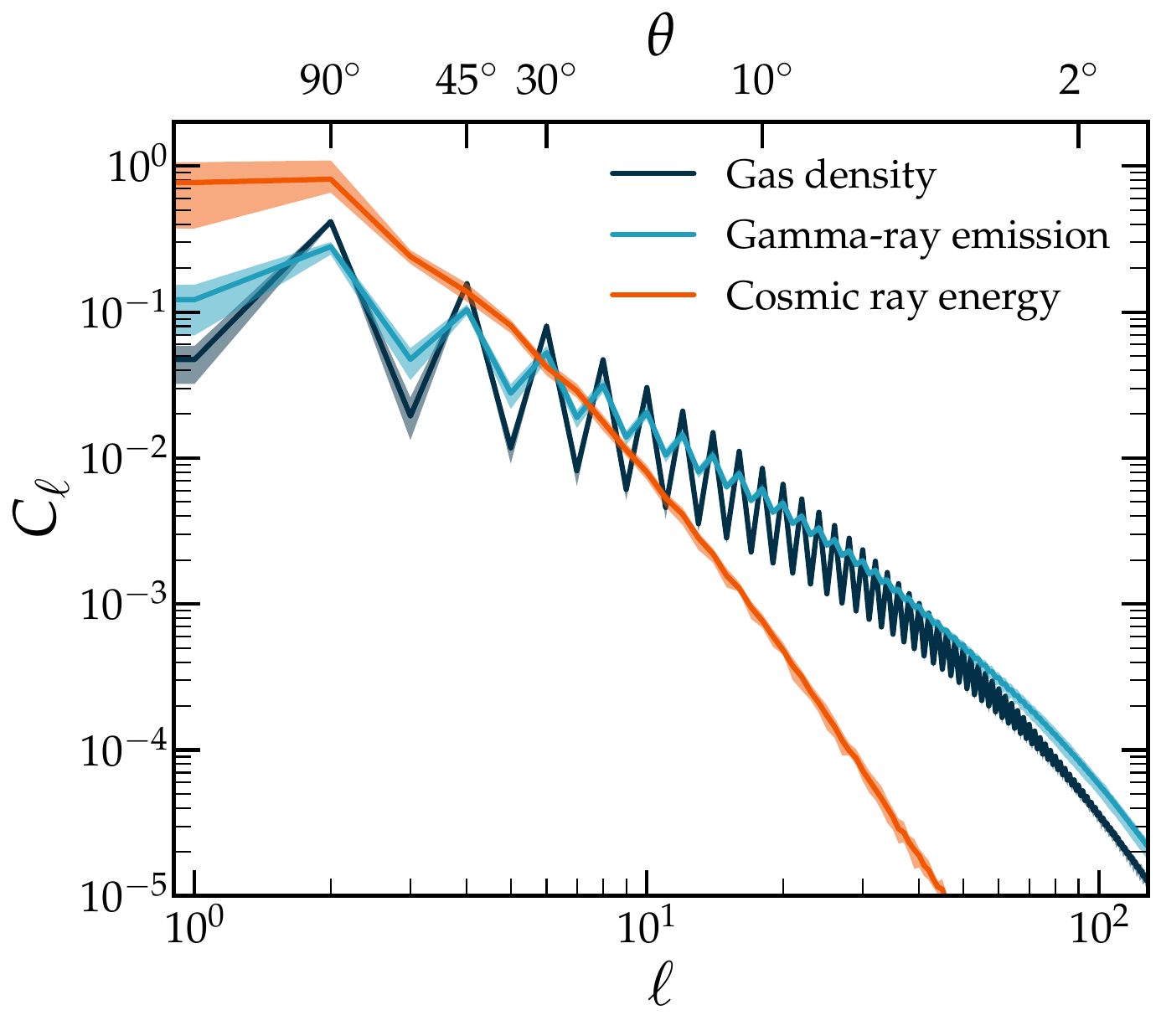}
    \caption{APS of the gamma-ray emission, column density, and projected CR energy as a function of multipoles $\ell$. The solid lines are medians of several bubbles, the shaded regions are the 25th and 75th percentiles. The saw-tooth pattern, visible in the density and gamma-ray emission, are an imprint of the galactic disk, since it is a large-scale anisotropic feature that is more dominant in the even multipoles due to its symmetry. The gamma-ray emission is mainly set by the gas density distribution.}
    \label{fig:aps}
\end{figure}

\begin{figure}
    \centering
    \includegraphics[width=\columnwidth]{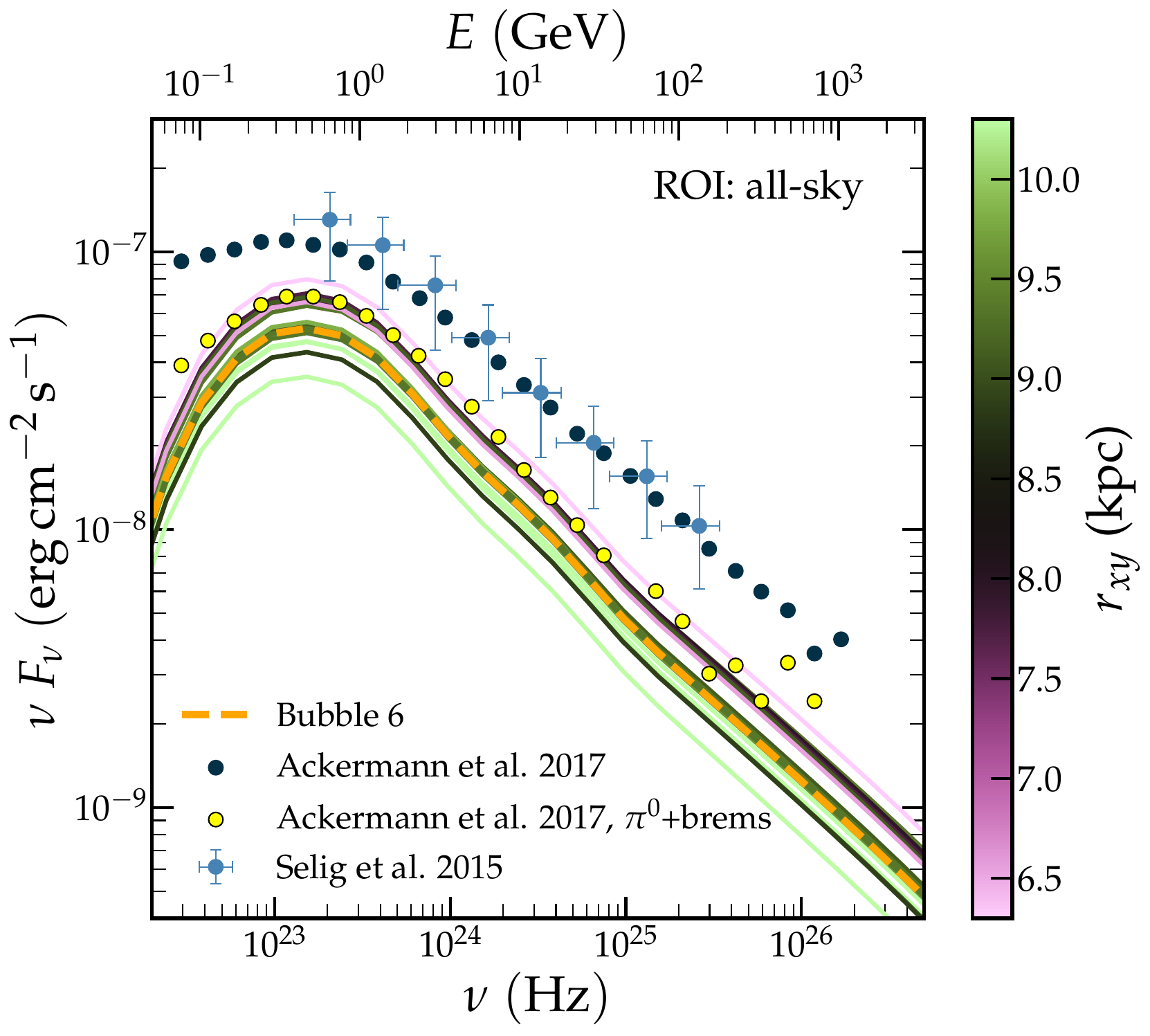}
    \caption{Comparing the all-sky gamma-ray spectra in simulations and observations. The colored lines are spectra from the perspective of different bubbles in the same snapshot; the dashed orange line is  bubble 6 in Fig. \ref{fig:bubbles}, which  we selected as our best candidate. The light and dark blue data points are observational results from \citet{Selig_2015} and \citet{Ackermann_2017}, respectively: both are based on 6.5 year Fermi-LAT data and show the total gamma-ray flux, \karin{with the point source contribution subtracted}. The yellow data points are specifically the diffuse gamma-ray emission from pion-decay and bremsstrahlung, based on GALPROP modeling \citep{Ackermann_2017}.}
    \label{fig:allsky}
\end{figure}

To zeroth order, the pion-decay gamma-ray emission is proportional to the product of the gas density and the CR energy density. To investigate whether the structure seen in the emission is preferentially determined by one or the other, we computed the angular power spectrum (APS) of the gamma-ray sky as well as the gas mass and CR energy projected onto the sky for a large number of bubbles, the results of which are shown in Fig.~\ref{fig:aps}. The APS tells us about the strength of fluctuations, i.e., the amount of structure, at different angular scales, represented by the multipole moment $\ell$, which scales inversely with angular size. Because the three fields differ substantially in their absolute normalization and dynamic range, we standardize each individual sky map prior to computing the APS. Specifically, for a given realization $i$ and field $X\in\{\mathrm{gamma,~gas,~CR}\}$ we compute $F^{(i)}_{X,\mathrm{stand}}=(F^{(i)}_X-\mu_{F^{(i)}_X})/\sigma_{F^{(i)}_X}$, where $\mu_{F^{(i)}_X}$ and $\sigma_{F^{(i)}_X}$ are the mean and standard deviation of the given map. The APS $C_\ell^{(i)}$ is then computed from $F^{(i)}_{X,\mathrm{stand}}$. The solid lines in Fig. \ref{fig:aps} show the average APS for each of the three fields, averaged over 21 bubbles, with the shaded regions showing the 20th and 80th percentiles. The narrow width of these bands indicates that there is little statistical difference between the bubbles, and that the structure is consistent between different observer locations. 

\begin{figure*}[htbp]
    \centering
    \includegraphics[width=\textwidth]{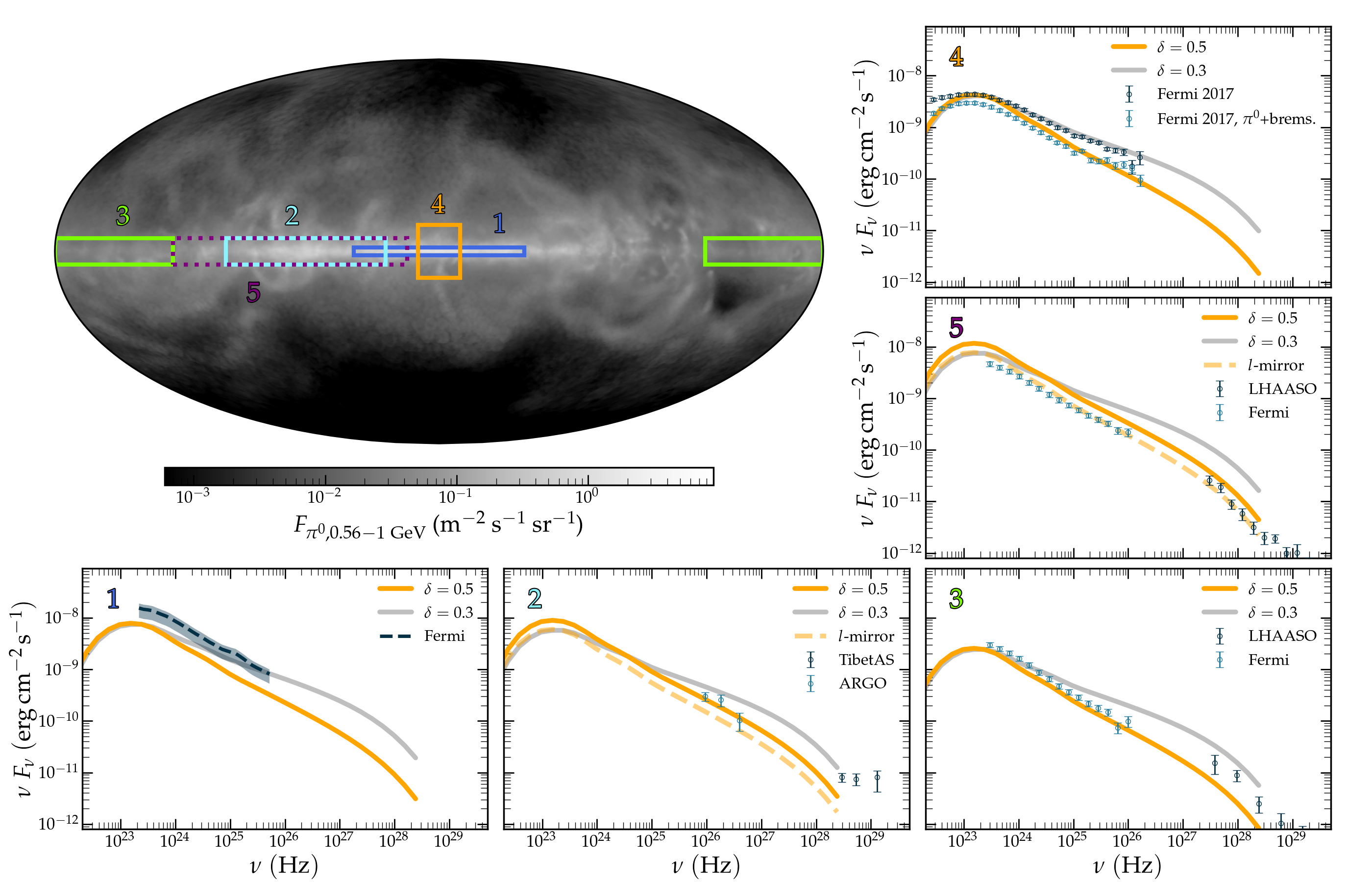}
    \caption{Comparison of gamma-ray fluxes to observations in different regions of the sky, for  bubble 6 in Fig. \ref{fig:bubbles}. The numbered boxes on the sky projection show which region is being counted, and the corresponding gamma-ray spectrum is shown in the smaller panels, marked by the same number. In orange is our fiducial run with $\delta=0.5$, while the gray line shows the spectrum with $\delta=0.3$. In the regions that are not symmetric in longitude (regions 2 and 5) the spectrum from the same region mirrored in longitude is shown. All other points and lines are from observations. The instruments used are given in the respective legends, and the data are  from \cite{Selig_2015} (region 1), \cite{bartoliSTUDYDIFFUSEGAMMARAY2015,amenomoriFirstDetectionSubPeV2021} (region 2), \cite{caoMeasurementUltraHighEnergyDiffuse2023,Zhang_2023} (regions 3 and 5), and \cite{Ackermann_2017} (region 4). We note that differences at the factor of $\sim$2 level may arise from both modeling details in the simulations and from analysis choices in the observations, such as point-source removal (see, e.g., Fig. A1 of \citet{Zhang_2023} for an illustration in region 5). Source confusion with diffuse emission can also affect the inferred spectral slope, particularly at low energies where confusion is more severe. We further note that, with the exception of region 4, the observations shown correspond to total gamma-ray emission rather than gas-correlated components. Regardless, we find remarkably good agreement between the simulations and observations.}
    \label{fig:fluxes}
\end{figure*}

The most eye-catching feature at first glance is the saw-tooth pattern, most evident in the gas density APS. This is, however, not a numerical issue, but an imprint of the galactic disk. So much of the total mass lies in a relatively thin strip around the galactic midplane that its structure is better captured in the even multipoles, and the power in the odd multipoles is suppressed. The CRs, in contrast, diffuse out of dense regions and tend to smooth out any sharp gradients, resulting in a much smoother APS with a steep decline in angular power toward larger multipoles.

The APS of the gamma-ray emission primarily follows the column density rather than the projected CR energy. As in the column density APS we see the imprint of the saw-tooth pattern associated with the midplane where most of the emission originates, although it is not as pronounced. Since the CR energy field is relatively diffuse, the structure seen in emission is set by the gas mass distribution. \karin{This is consistent with early GALPROP modeling, which found that in order to reproduce the gamma-ray emission large CR halo heights of several kiloparsec are needed \citep{strongPropagationCosmicRayNucleons1998}, and that varying the size of the halo between 2 and 10 kpc had little effect on the total luminosity.}

\section{Comparison to observations}\label{sec:observations}
While we found that our luminosities are consistent with the Milky Way (see Fig.~\ref{fig:fcal}), it is of interest to see whether this also applies to the gamma-ray sky as seen by an observer, which we can directly compare to observed fluxes of the Galaxy. The all-sky flux spectra for 18 simulated bubbles are shown together in Fig.~\ref{fig:allsky}, colored by the galactocentric radius $r_{xy}$ of the observer, together with the observed spectrum of diffuse gamma rays based on 6.5-year Fermi-LAT data \citep{Selig_2015,Ackermann_2017}. In addition to the total measured spectrum, \citet{Ackermann_2017} decomposed the diffuse emission into individual components using GALPROP modeling, and we therefore also show their inferred contribution from pion decay and bremsstrahlung. Our simulated all-sky flux lies a factor of $\sim$2–3 below the total observed \karin{diffuse} gamma-ray emission, but is fully consistent with the expected hadronic component in both normalization and spectral slope. \karin{To capture the full observed flux we would also have to  include the leptonic gamma-ray emission as well as the isotropic background flux}. The normalization of the all-sky spectrum does vary slightly between different observers, which roughly correlates with galactocentric radius.

In addition to the all-sky spectrum, we computed the flux spectrum for different regions of the sky that have been studied observationally. These regions are shown and labeled in Fig.~\ref{fig:fluxes}: $|b|<1.5\degree,|l|<40\degree$ (region 1) based on observations by Fermi-LAT \citep{Selig_2015}; $|b|<5\degree,l=[25\degree,100\degree]$ (region 2) based on observations by ARGO \citep{bartoliSTUDYDIFFUSEGAMMARAY2015} and TibetAS \citep{amenomoriFirstDetectionSubPeV2021}; $|b|<5\degree,l=[15\degree,125\degree]$ (region 5) and $l=[125\degree,235\degree]$ (region 3) based on observations by LHAASO \citep{caoMeasurementUltraHighEnergyDiffuse2023} and Fermi-LAT \citep{Zhang_2023}; and $|b|<10\degree,|l|<10\degree$ (region 4) based on observations by Fermi-LAT \citep{Ackermann_2017}. In the case of regions 2 and 5, which are not symmetric in longitude, we additionally show the spectra from the corresponding mirrored regions, since this is an arbitrary choice in the simulation. 

\begin{figure*}
    \centering
    \includegraphics[width=\textwidth]{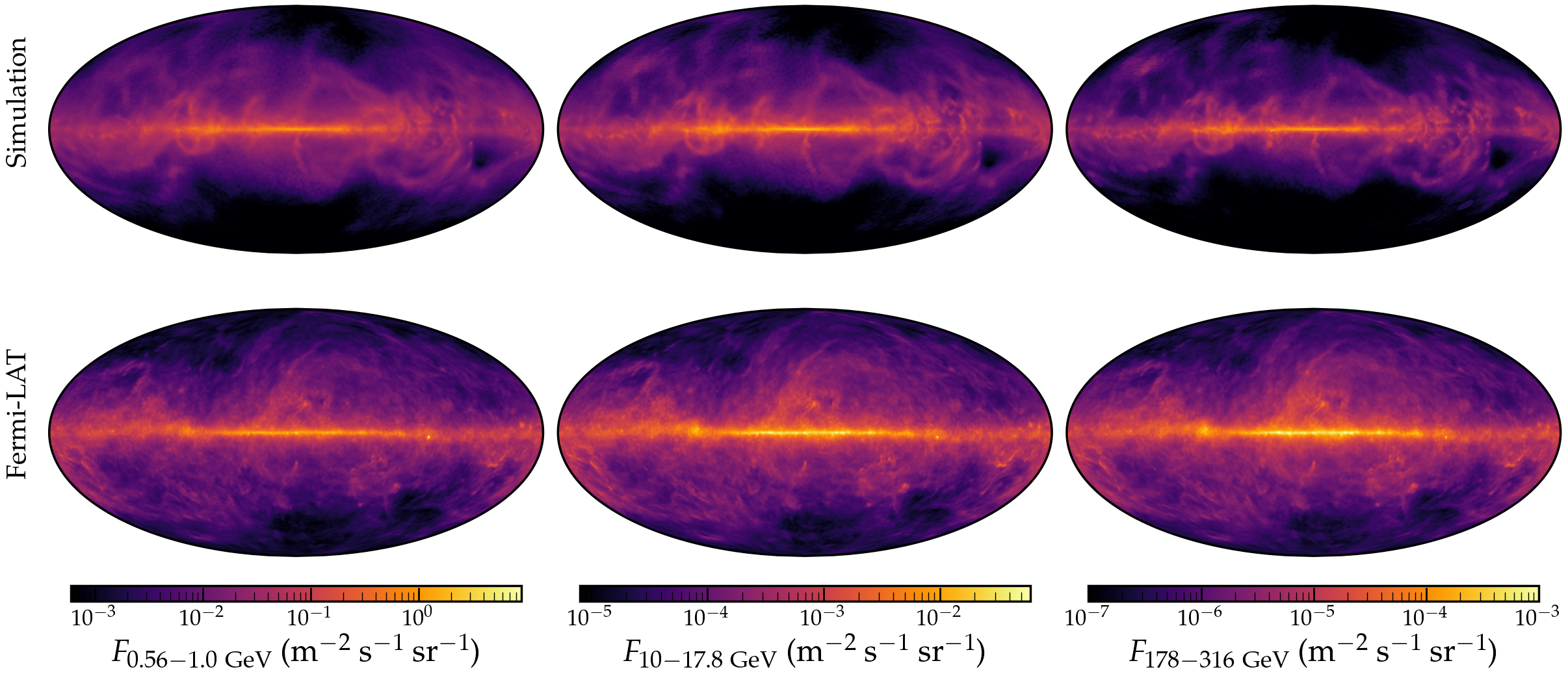}
    \caption{Comparison of the diffuse gamma-ray emission between our simulations (upper panels) and observations (lower panels) by Fermi-LAT \citep{Scheel-Platz_2023}, in three different energy bands: 0.56-1.0 GeV (left), 10-17.8 GeV (middle), and 178-316 GeV (right). The top row shows the gamma-ray sky for the same bubble as in Fig.~\ref{fig:d90_flux}. The bottom row shows the dust-correlated component of the observed diffuse gamma-ray emission.}
    \label{fig:fermi_comparison}
\end{figure*}

For most regions, our fiducial model with $\delta=0.5$ reproduces the observed spectra well. In particular, regions~3 and~5 show a striking agreement between the simulated flux and the Fermi-LAT measurements. However, because the observational data include both hadronic and leptonic contributions, whereas our simulations only model hadronic emission, such a close agreement \karin{could} indicate a mild overestimation of the hadronic flux in these regions. Region 4 provides a useful benchmark since \cite{Ackermann_2017} provide estimates of the pion-decay and bremsstrahlung component, which  lies a factor of $\lesssim 2$ below the total observed diffuse emission. Ideally, our simulated flux in the other regions should therefore also fall below the observed total emission by a similar amount. However, it should not be overinterpreted as a discrepancy in the regions where it does not. In regions 2 and 5 we can get a comparable variation in the normalization by simply considering the mirrored region across the galactic center. \karin{We also need to be mindful of the possibility of point source confusion in the observational data. For example, for the data shown in regions 3 and 5, the same point source masking was used by both \cite{Zhang_2023} and \cite{caoMeasurementUltraHighEnergyDiffuse2023}, based on LHAASO sources. \cite{Zhang_2023} note, however, that this likely removes a nonnegligible fraction of the diffuse emission, as the flux is noticeably reduced in the inner region (region 5) compared to when subtracting point sources using the Fermi catalog.}

The Galactic Center is a complicated region with puzzling gamma-ray measurements that could possibly hint at unresolved millisecond pulsars \citep{Abazajian2011, AbazajianKaplinghat2012, YuanBing2014} or dark matter annihilation \citep{huang2016,Daylan_2016,Ackermann_2017}. In the simulations we are also missing certain features of the Galactic Center, such as a galactic bar and the Fermi bubbles. However, despite this, our simulation reproduces the observed flux from this region well. In region 1 the simulated flux falls below the Fermi-LAT measurements from \citet{Selig_2015}, as expected given that these data include diffuse emission from channels beyond pion decay. In region 4, which targets a similar area, the data also mostly lie below the total observed emission, but it seems we have overestimate the amount of flux near the spectral peak when we make a comparison only to the hadronic+bremsstrahlung component.

Assuming a scaling of $\delta=0.5$ with energy for the CR diffusion coefficient changes the high-energy tail of the spectra, and generally fits the observational data better. There are regions, however, where the slope seems better captured by a weaker scaling (region 4) or where it is unclear which scaling is better. We discuss the choice of $\delta$ in Section~\ref{sec:disc}.

\begin{figure*}
    \centering
    \includegraphics[width=\textwidth]{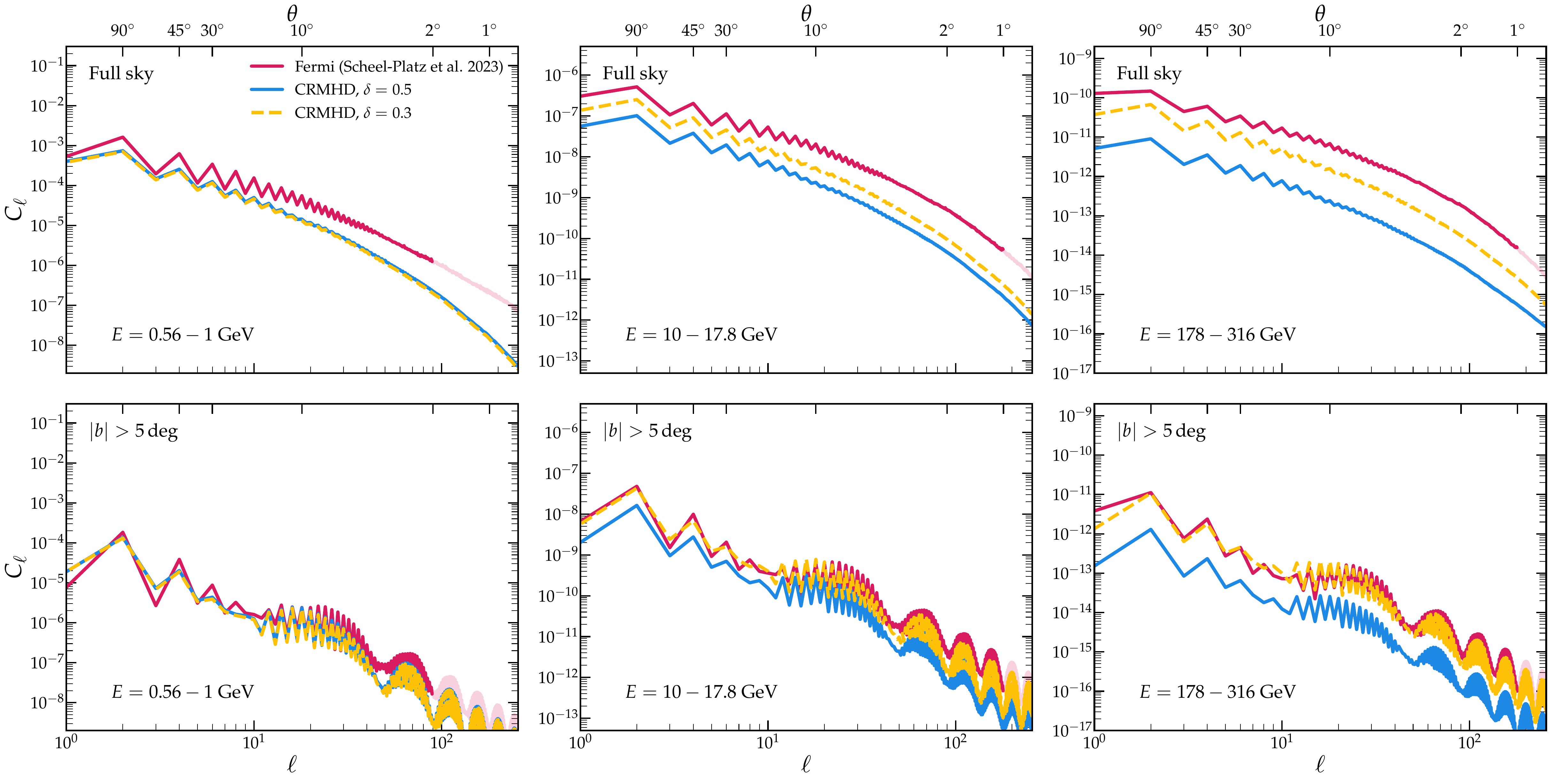}
    \caption{APS of the maps shown in Fig. \ref{fig:fermi_comparison}. Top row: APS of the full gamma-ray sky in different energy bands. Bottom row: APS of the sky, omitting all pixels with $|b|<5\degree$. In pink are the observations, \karin{where we limit the APS to $\theta>2^\circ$ in the low-energy bin and $\theta>1^\circ$ in the two higher-energy bins, due to the PSF mismodeling artifacts present in the reconstruction}. In blue and yellow are the APS from our simulated gamma-ray sky with $\delta=0.5$ and $\delta=0.3$ respectively.}
    \label{fig:APS_comp}
\end{figure*}

Figure~\ref{fig:fermi_comparison} compares the diffuse gamma-ray emission in our simulation with the reconstructions of \cite{Scheel-Platz_2023}, who used ten-year Fermi-LAT data and a Bayesian inference framework to decompose the diffuse gamma-ray emission into two components. The one shown here (bottom row) is the dust-correlated diffuse emission, i.e., the diffuse emission whose target population distribution traces the gas density. This should therefore predominantly target the hadronic component of the emission, which is what we simulate, as well as bremsstrahlung emission. It is important to note, however, that the relationship between dust emission and gas column density is not straightforward \citep[e.g.,][]{renCosmicRaysGas2025}. \karin{Opacity steeply rises in denser clouds as you move into the molecular phase \citep{remy_cosmic_2017}}, meaning that some caution is required when equating dust-correlated emission with CR proton emission. 

Quantitatively there is good agreement between the maps in all three energy bins, and the dynamic ranges are similar. There is small-scale structure evident in all maps, though it is more prominent in the observations. In the two higher-energy bins it looks as if we are missing some flux at higher latitudes and toward the galactic center, in particular in the $178-316$ GeV bin. This is somewhat puzzling considering the simulated all-sky flux spectra in Fig. \ref{fig:allsky} and the GALPROP predictions (yellow data points), which show especially good agreement in the energy range $0.1-100~\mathrm{GeV}$. \karin{Because the relationship between dust opacity and column density becomes nonlinear at high densities, correlations between dust emission and column density may overestimate the gas content, which might bias the diffuse emission in the reconstruction toward higher values, in particular in the inner Galaxy. \cite{Scheel-Platz_2023} also acknowledge that there is a mismatch in the point spread function (PSF) modeling, resulting in possible point-source contamination (which typically have harder spectra than diffuse emission) as well as an artificial sharpening of the emission from the disk. Together, these factors could partially explain the differences we see between our simulation and the reconstruction in the higher-energy bins.}  

Figure~\ref{fig:APS_comp} compares the APS of the diffuse gamma-ray skies shown in Fig.~\ref{fig:fermi_comparison}, as well as the simulation with $\delta=0.3$. The top row shows the APS applied on the full gamma-ray sky, while the bottom row shows the APS for the sky with $b>5\degree$, i.e., omitting the galactic midplane. 

\begin{figure}
    \centering
    \includegraphics[width=\columnwidth]{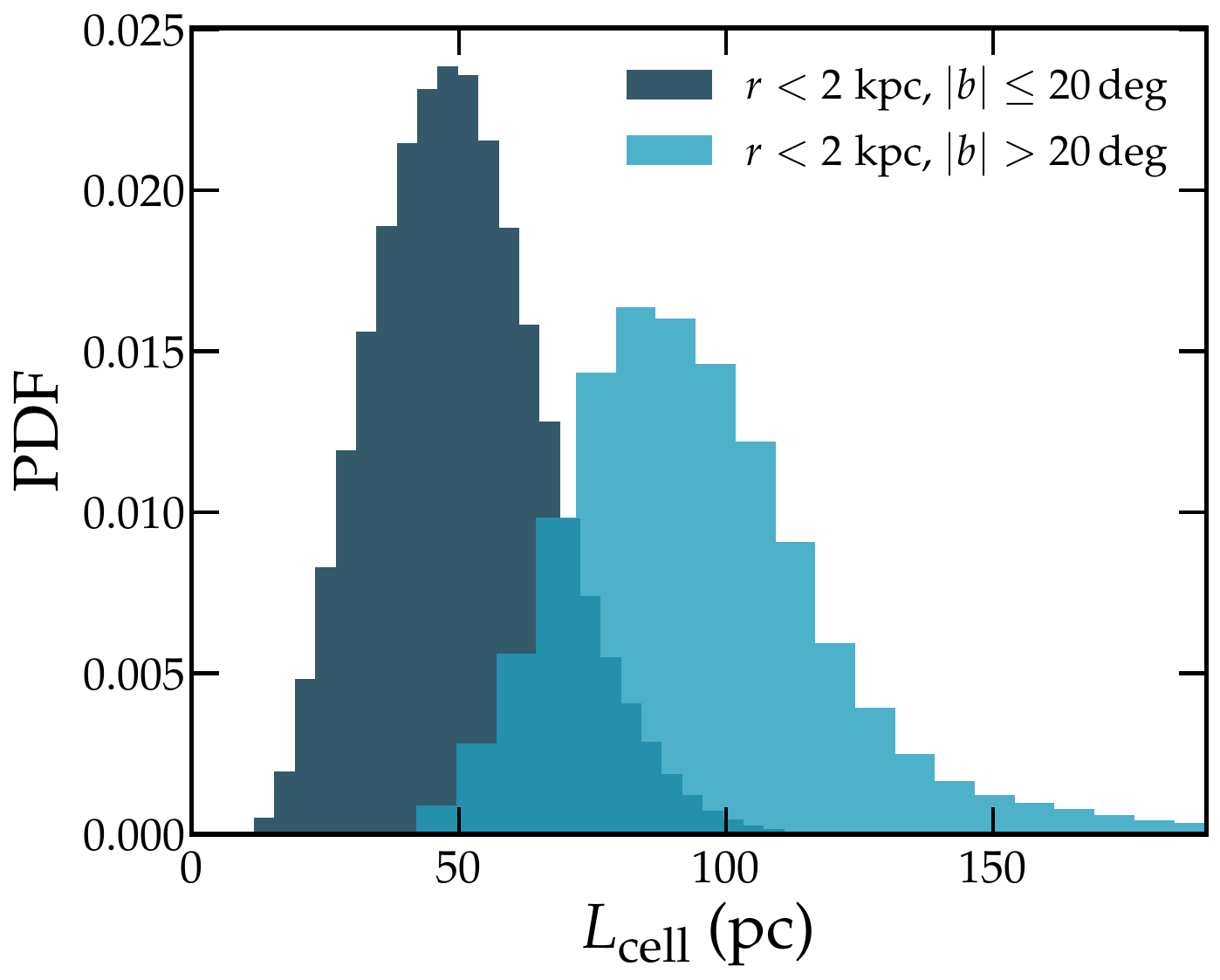}
    \caption{Histogram of the cell sizes of all the cells within 2 kpc of an observer, averaged over 21 different bubbles and weighted by the gamma-ray luminosity. In dark blue are cells with latitudes $|b|<20\degree$, and in light blue  cells with $|b|>20\degree$.  }
    \label{fig:pdf}
\end{figure}

In the lowest energy bin (0.56--1 GeV), the all-sky APS shows good agreement between simulations and observations on large angular scales (low $\ell$), but the two diverge at higher multipoles, where the observations exhibit a greater amount of small-scale structure. \karin{In the \cite{Scheel-Platz_2023} reconstruction there is an artificial amount of power at the smallest scales due to the PSF modeling mismatch, for this reason we limit the APS to $\theta>2^\circ$ in the lowest-energy bin and $\theta>1^\circ$ in the two higher-energy bins. Nonetheless, our simulation is likely underestimating the small-scale power due to a} combination of limited simulation resolution and simplifications in our CR modeling. Finite resolution restricts the smallest spatial scales that can be represented and can therefore suppress power at large $\ell$, particularly for emission originating at nearby distances. This is illustrated in Fig. \ref{fig:pdf}, which shows a histogram of cell sizes within 2 kpc of the simulated observer, averaged over 21 realizations and weighted by gamma-ray luminosity. As seen in Fig. \ref{fig:d90_b}, for galactic latitudes $|b|>20\degree$ the sky is dominated by emission from cells $\leq2$~kpc away, which have sizes of $\sim$75-100 pc (light blue histogram). As an example, a cell of transverse size $x=75~\mathrm{pc}$ at a distance $d=2~\mathrm{kpc}$ away, approximately corresponds to an $\ell$-mode of $\ell=\pi/(x/d)\approx84$. Considering this, it is not unreasonable that we could be suppressing power for large multipoles, which emphasizes the need for higher-resolution simulations.  

In addition, with the steady-state approximation applied to the simulated gray CRs, there is little spectral variation across the sky. In reality, low-energy CRs lose their energy faster \citep[e.g.,][]{evoliAMS02BerylliumData2020} and therefore trace the regions where they are injected, while the higher-energy CRs propagate farther and produce a smoother distribution. This energy dependence could enhance small-scale structure at lower energies, an effect that is not fully captured in our modeling.

In the two higher-energy bins the shape of the full-sky APS is well reproduced in the simulation, but the overall normalization is systematically lower than observed. For the $\delta=0.5$ model, this discrepancy increases with energy. Using a weaker scaling of the diffusion coefficient in this case does naturally increase the intensity at higher energy, but we still do not capture the same amount of power as in the observations. By masking away the midplane, shown in the bottom row of Fig. \ref{fig:APS_comp}, we ignore any potentially unresolved point sources that could be present in the reconstructed sky-maps, and by doing this we achieve a close match between the observations and the $\delta=0.3$ model. This could suggest that a weaker energy dependence of the diffusion coefficient is more appropriate at higher latitudes, and once again emphasizes the importance of accurately modeling CR transport when interpreting the details of the gamma-ray emission.

\section{Discussion}\label{sec:disc}
\subsection{CR transport details}
The choice of CR transport parameters has a clear impact on the resulting gamma-ray spectrum. Our simulations assume one constant diffusion coefficient throughout the Galaxy, while our steady-state modeling of the CR proton spectrum assumes one constant value of the energy dependence $\delta$ of the diffusion coefficient. In the case of external confinement of CRs, where their scattering is dominated by externally driven turbulence, a scaling of $\delta=0.3$ is expected for isotropic Kolmogorov turbulence \citep{ruszkowskiCosmicRayFeedback2023}. For this work we  used $\delta=0.5$ based on work by \cite{evoliAMS02BerylliumData2020} who find a slope of $\delta=0.54$ by fitting experimental data of secondary to primary ratios by AMS-02. In other works an even steeper slope is found \citep{maurinSystematicUncertaintiesCosmicray2010}. Our results favor $\delta=0.5$ rather than $\delta=0.3$, but this should not be overinterpreted. Most modeling of the energy dependence of CR diffusion assumes a spatially constant diffusion coefficient that is only a function of particle momentum, but the diffusive properties of CRs likely change in different regions of the Galaxy. The nature of turbulence is different in the halo compared to the disk, which  causes a latitudinal change in $\delta$ in the case where CRs scatter off externally driven turbulence \citep{tomassettiOriginCosmicRay2012}. 
 
Previous simulation work has shown how the effective diffusion coefficient (for GeV CRs) varies in different phases of the ISM if CR streaming is accounted for \citep{armillottaCosmicRayTransportVarying2022,thomasCosmicraydrivenGalacticWinds2023,thomas2025_effective}. In this case the CRs excite Alfvén waves through gyroresonance, which are damped by  processes such as ion-neutral or nonlinear Landau damping \citep{xu_cosmic_2022,Thomas2025}. Differences in the effectiveness of the damping processes, or in which process dominates,  affects the diffusion coefficient. This would also affect the resulting gamma-ray spectrum. Considering how well our results fit with the observational data, we expect this effect to be minor. 

We  only accounted for diffusion parallel to the magnetic field lines, when in reality a small amount of perpendicular diffusion is expected. The degree of anisotropy is not known, but is generally estimated to be $D_\perp/D_{||}=10^{-4}-10^{-1}$ \citep{shalchi2010}, and could depend on energy \citep{dundovicNovelAspectsCosmic2020}. \cite{dornerTestingInfluenceAnisotropic2025} tested the effect of varying the degree of anisotropy on the gamma-ray sky, and found only small changes in the flux and the overall tomography of the sky.

\subsection{Leptonic emission}
For this work, we  modeled the gamma-ray emission from neutral pion decay, and found good agreement between our simulations and the observations. However, we did not include the contribution of CR electrons to the emission, which warrants discussion. In the Milky Way, the gamma-ray emission from neutral pion decay is believed to dominate in the 0.1--100 GeV range, with leptonic processes contributing 20--40\% for typical values of the electron-to-proton ratio of $\sim$0.01 at GeV energies \citep{strongGLOBALCOSMICRAYRELATEDLUMINOSITY2010,martinInterstellarGammarayEmission2014,Werhahn_2021_II}. Roughly 15--30\% of the total luminosity is contributed by IC, and 7--20\% from bremsstrahlung. IC  dominates the emission at $\lesssim$$10~\rm MeV$, and is also  important at the highest energies, depending on the interstellar radiation field (ISRF). Bremsstrahlung likely becomes more important at lower energies ($\lesssim$$10~\rm GeV$), though this  depends on the gas density \citep{martinInterstellarGammarayEmission2014}. However, the exact contribution of IC emission, in particular, is  highly dependent on the model of electron injection and propagation, and is limited by the uncertainties involved in the spatially varying ISRF; it is often quoted as being one of the largest sources of uncertainty in modeling of the diffuse gamma-ray emission in the Milky Way \citep[e.g.,][]{ackermannSPECTRUMISOTROPICDIFFUSE2015}. This further justifies our choice of neglecting this component for this particular work.

Like the hadronic emission, bremsstrahlung is correlated with the gas, so its inclusion is unlikely to affect our conclusions regarding the role of local emission at higher Galactic latitudes. In contrast, IC emission might be more diffuse and will subsequently have a much smoother distribution, and is therefore  more important above and below the mid-plane \citep{Selig_2015,Werhahn_2021_II,Scheel-Platz_2023}, while the density-correlated components always dominate near the Galactic plane at gigaelectronvolt energies. Though we believe our findings to be robust, it will be interesting to see how the inclusion of IC emission will impact our results.

\section{Conclusions}\label{sec:conclusions}
In this work we  simulated and analyzed gamma-ray emission from pion decay of simulated, isolated, Milky Way-like galaxies from the Rhea suite \citep{goeller2025,kjellgren2025}. The simulations are run using the moving-mesh code \textsc{arepo}, and evolve the total integrated CR proton energy density in a self-consistent manner. The gamma-ray emission from neutral pion decay is calculated in post-processing using \textsc{crayon+} \citep{Werhahn_2021_I,Werhahn_2021_II}, which accounts for the various energy-dependent losses of CR protons to calculate a steady-state spectrum in each computational cell. This is done without fine-tuning the source spectrum to match observations, and provides a self-consistent calculation of the emission based on the parameters of our galaxy simulation and a few parameter assumptions, such as the energy-dependence of the diffusion coefficient. We used this to investigate the impact of local emission on the full gamma-ray sky seen by different observers, \karin{and succeeded in reproducing several well-established observational trends of Galactic gamma-ray emission.}

Our main conclusions are as follows:

\begin{enumerate}
    \item We successfully reproduced the gamma-ray luminosity expected for the Milky Way in our simulations. The total luminosity varies moderately with time due to the time variability of the SFR, meaning that it is possible to find a simulation time that closely matches the modeled luminosity spectrum of the Milky Way. The energy scaling of the diffusion coefficient is important, with models using $\delta=0.3$ overestimating the luminosity at higher energies ($>$10 GeV). A scaling of $\delta=0.5$ reproduces the correct luminosity spectrum significantly better, showing how sensitive our results are to the details of CR transport.
    \item The gamma-ray sky an observer sees can vary drastically for different positions in the galaxy, thus showing that the local environment is important. In particular, at high latitudes the local arms and spurs differ perceptibly between bubbles. 
    \item For a given gamma-ray sky seen by an observer, it is common for lines of sight with prominent features and/or filaments in the gamma-ray emission to come from the local ($\lesssim$2 kpc) environment. Local emission becomes increasingly more important with increasing galactic latitude.
    \item The structure of the gamma-ray emission in the sky is primarily set by the gas density distribution rather than the CR energy density distribution, which is much more diffuse. 
    \item We reproduced the all-sky flux spectrum from pion-decay, which varies slightly depending on which bubble the observer is placed in. The observed gamma-ray flux from different regions of the sky is consistent with observations. As is the luminosity, the emission spectrum is sensitive to the choice of CR transport parameters, and in most cases $\delta=0.5$ is favored. 
\end{enumerate}

\begin{acknowledgements}
\karin{We thank the anonymous referee for a careful reading of the manuscript and the many insightful and valuable comments that helped improve the quality of the paper.} The team in Heidelberg  acknowledges financial support from the European Research Council via the ERC Synergy Grant ``ECOGAL'' (project ID 855130),  from the German Excellence Strategy via the Heidelberg Cluster of Excellence (EXC 2181 - 390900948) ``STRUCTURES'', and from the German Ministry for Economic Affairs and Climate Action in project ``MAINN'' (funding ID 50002206).00
The authors gratefully acknowledge the scientific support and HPC resources provided by the Erlangen National High Performance Computing Center (NHR@FAU) of the Friedrich-Alexander-Universität Erlangen-Nürnberg (FAU) under the NHR project a104bc. NHR funding is provided by federal and Bavarian state authorities. NHR@FAU hardware is partially funded by the German Research Foundation (DFG) – 440719683.
They also thank for computing resources provided by the Ministry of Science, Research and the Arts (MWK) of the State of Baden-W\"{u}rttemberg through bwHPC and the German Science Foundation (DFG) through grants INST 35/1134-1 FUGG and 35/1597-1 FUGG, and for data storage at SDS@hd funded through grants INST 35/1314-1 FUGG and INST 35/1503-1 FUGG.
NB acknowledges support from the ANR BRIDGES grant (ANR-23-CE31-0005).
KK is a fellow of the International Max Planck Research School for Astronomy and Cosmic Physics at the University of Heidelberg (IMPRS-HD). CP acknowledges support from the European Research Council via the ERC Advanced Grant ``PICOGAL'' (project ID 101019746).
\end{acknowledgements}

\bibliographystyle{aa} 
\bibliography{references}

\begin{appendix}

\section{Validity of steady-state assumption}\label{app:steadystate}

\begin{figure}
    \centering
    \includegraphics[width=\columnwidth]{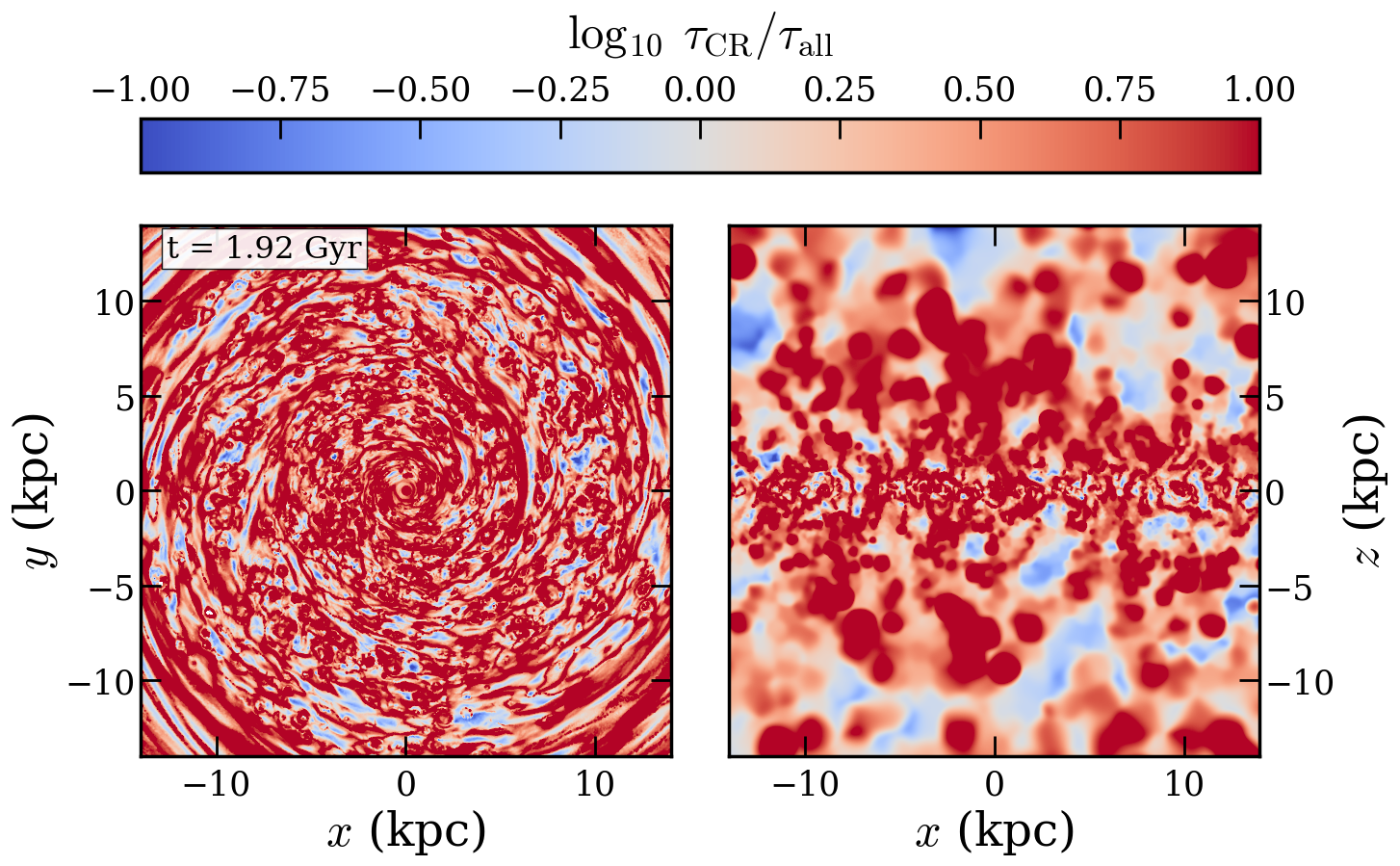}
    \vfill
    \includegraphics[width=0.7\columnwidth]{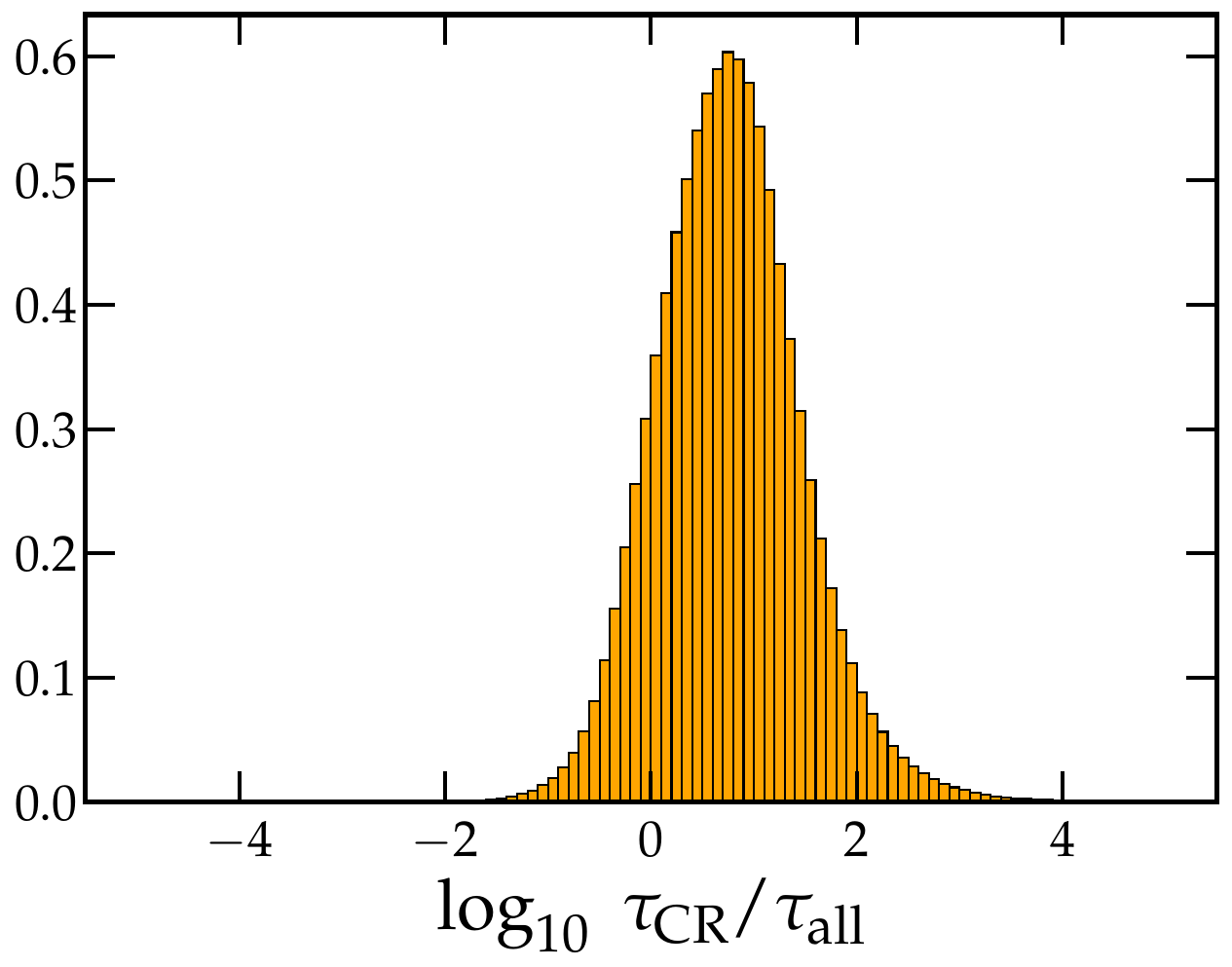}
    \caption{Top: face-on and edge-on slice through the midplane showing the ratio of the timescale at which the CR energy density changes $\tau_{\rm CR}$, to the timescale of losses and escape $\tau_{\rm all}$, evaluated in every computational cell. Bottom: mass-weighted histogram of the same ratio in each cell.}
    \label{fig:steadystate}
\end{figure}

\karin{The steady-state approximation relies on the assumption that the timescale at which CR losses and escape occur, defined as $\tau^{-1}_{\rm all}=\tau^{-1}_{\rm esc}+\tau^{-1}_{\rm cool}$, is shorter than the timescale over which the CR energy density changes, $\tau_{\rm CR}$. We evaluate this assumption in Fig.~\ref{fig:steadystate} where we show in the left panel the ratio of these two timescales $\tau_{\rm CR}/\tau_{\rm all}$ in a face-on and edge-on slice through the center of our galaxy, and in the right panel where we show a mass-weighted histogram of each cell. The escape timescale $\tau_{\rm esc}$ is defined as a combination of diffusion and advection, $\tau^{-1}_{\rm esc}=\tau^{-1}_{\rm diff}+\tau^{-1}_{\rm adv}$, where the advection and diffusion timescales are defined in Eq.~\eqref{eq:t_esc}. The cooling timescale is defined as the total CR energy divided by the sum of the hadronic and Coulomb energy losses, $\tau_{\rm cool}=e_{\rm CR}/|b_{\rm hadr}+b_{\rm Coul}|$. The timescale at which the CR energy density in each cell changes is calculated as $\tau_{\rm CR}=e_{\rm CR}/(\Delta e_{\rm CR}/\Delta t)$, where $\Delta e_{\rm CR}$ is change in CR energy density between two consecutive snapshots separated by $\Delta t$. Though there are a few regions where the steady-state assumption breaks down, the vast majority of cells obey that $\tau_{\rm CR}\gtrsim\tau_{\rm all}$, owing to fast hadronic and diffusive losses. This result agrees with the steady-state analysis using a smooth pressurized ISM \citep[figure~9 of][]{Werhahn_2021_I} and an analysis of the simulated radio-synchrotron emission in an edge-on galaxy \citep[figure~A1 of][]{chiuQiuSimulatingRadioSynchrotron2024}, adopting the structured multi-phase ISM model CRISP \citep{Thomas2025}, similar to the present study.}

\section{Comparison between bubbles}\label{sec:app1}
\begin{figure}
    \includegraphics[width=\columnwidth]{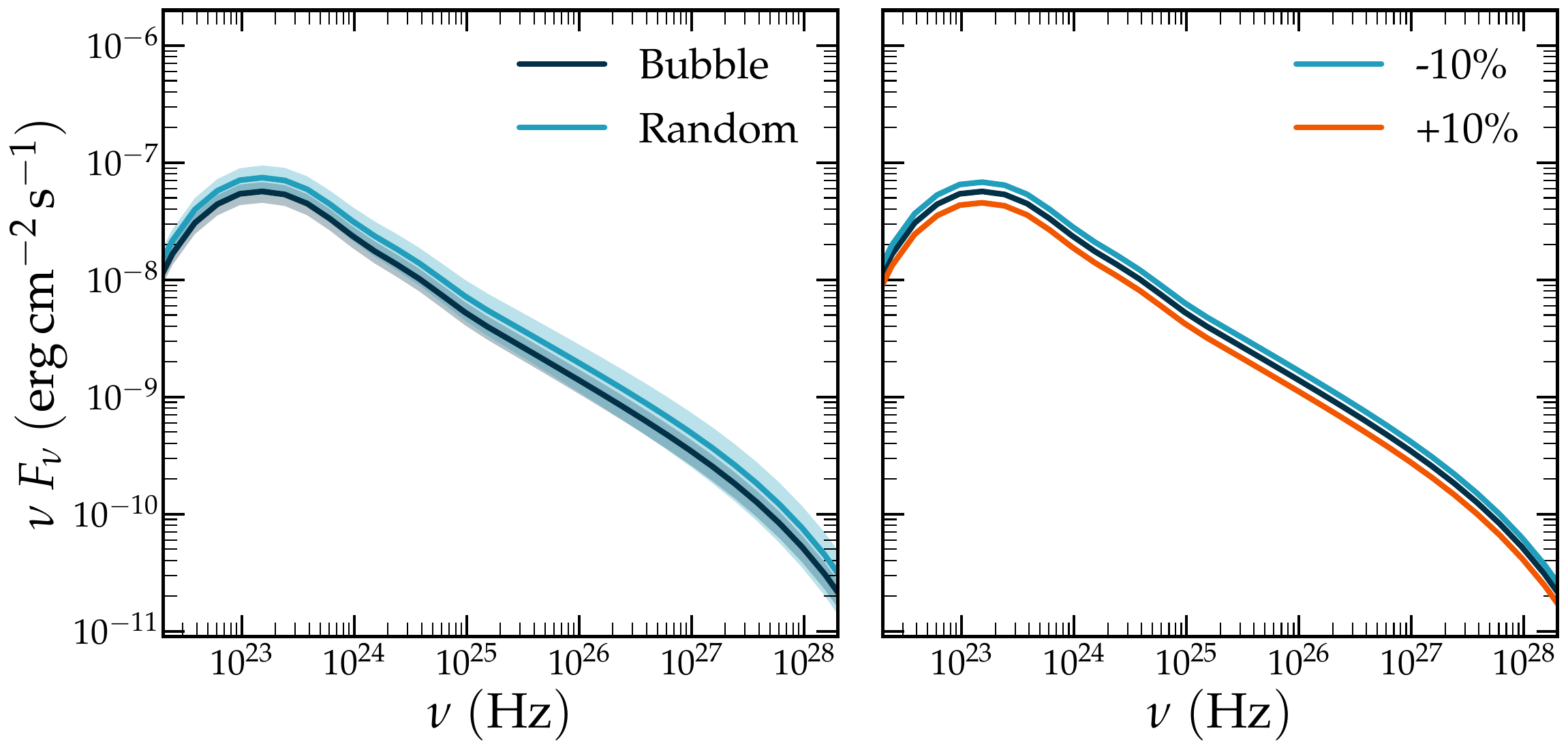}
    \caption{Left: average all-sky flux spectrum from 18 bubbles compared to 18 randomly selected positions at similar galactocentric radii. Shaded regions are $1\sigma$ uncertainty intervals. Right: Average flux spectrum of 18 bubbles together with the average spectra if there was a $\pm10\%$ uncertainty in the distances to the flux-emitting cells.}
    \label{fig:fullspeccomp}
\end{figure}

For completeness's sake, and to show the variation between different observer locations, we include some analysis on other bubbles. All bubbles were selected by taking a density-slice through the midplane and picking out locations in low-density regions, with galactocentric radii $r_{xy}$ within 2 kpc of the solar circle. Though we found that there are only minor variations between different bubbles, we can ask ourselves how much of an impact the specific bubble location has. The left panel of \ref{fig:fullspeccomp} shows the average all-sky spectrum for a selection of 18 hand-picked bubbles versus the same number of random locations in the disk. We see that while the flux from random locations is slightly higher, the variations are minor. The right panel shows how much the flux spectrum changes if we assume the distances to the emitting cells have an uncertainty of $\pm10\%$ \citep{hunterTestingKinematicDistances2024}. Using standard error propagation, one can analytically show that this corresponds to an uncertainty of $\pm20\%$ in the flux, in agreement with our results.

\begin{figure*}[htbp]
  \centering
  \includegraphics[width=0.49\textwidth]{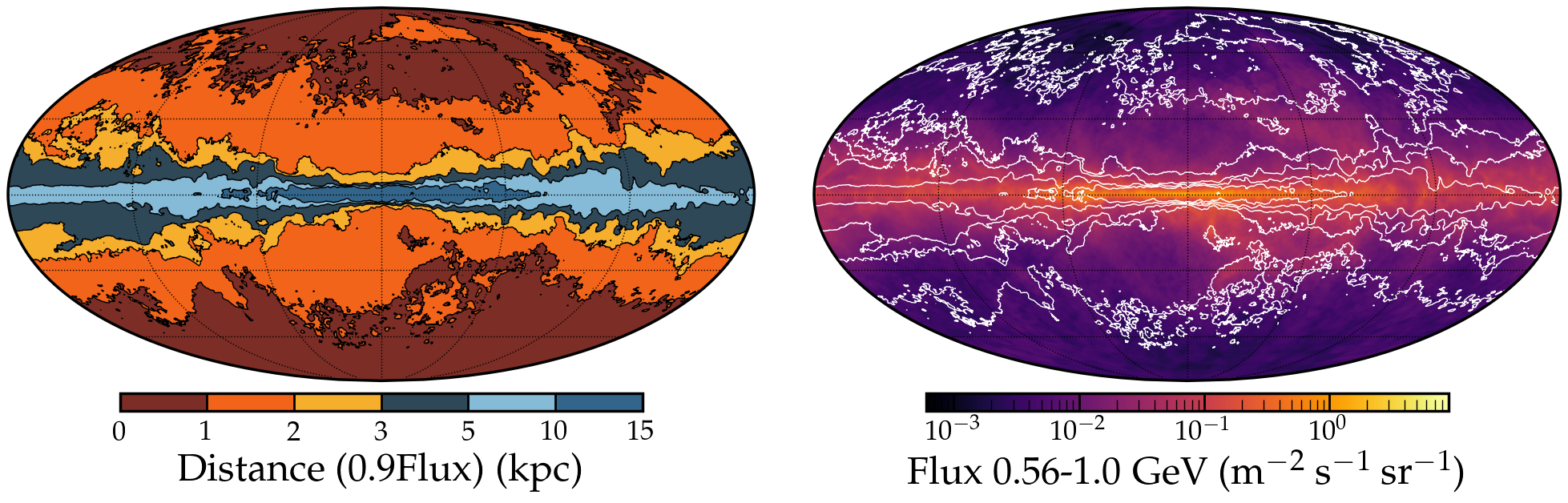}
  \hfill
  \includegraphics[width=0.49\textwidth]{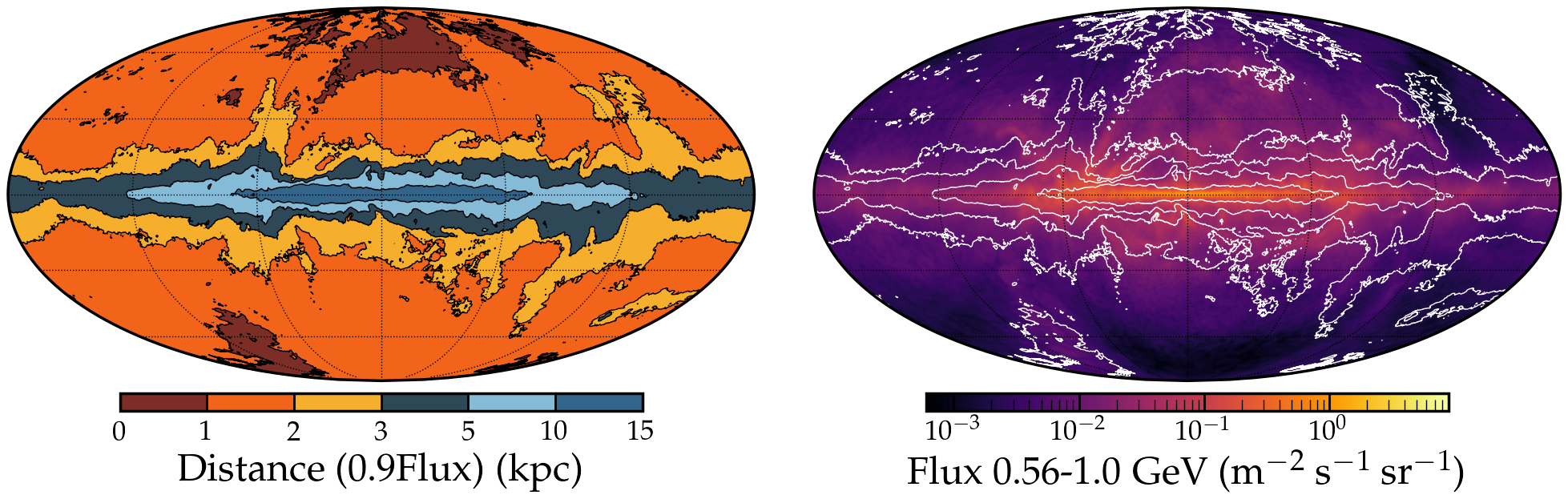}
  \includegraphics[width=0.49\textwidth]{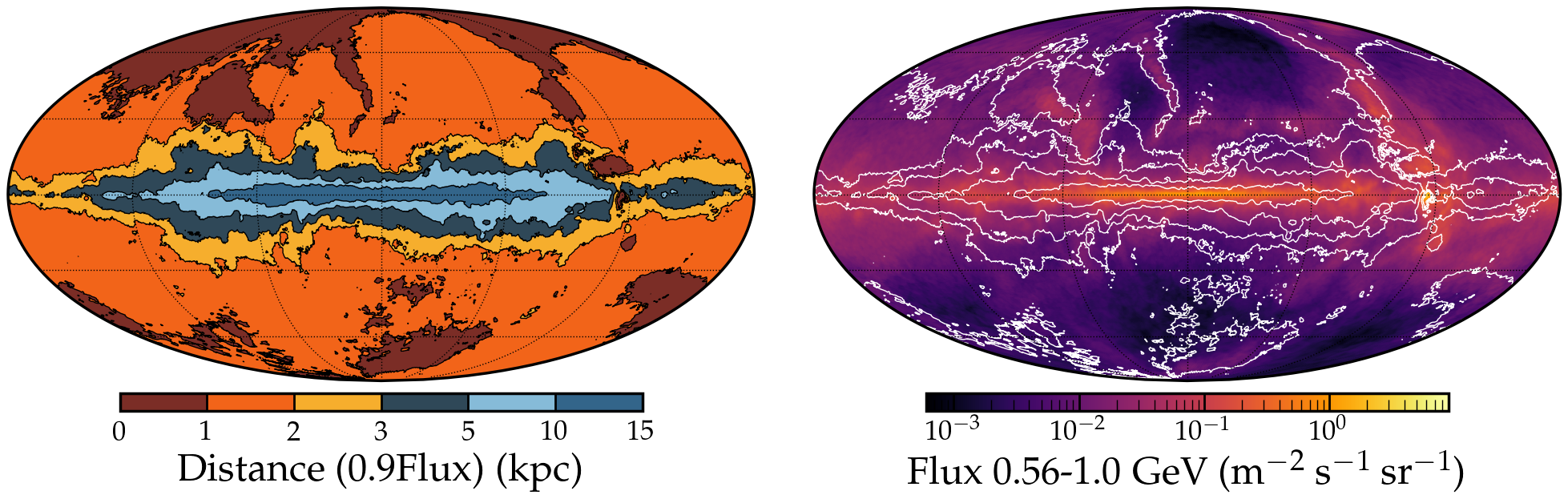}
  \hfill
  \includegraphics[width=0.49\textwidth]{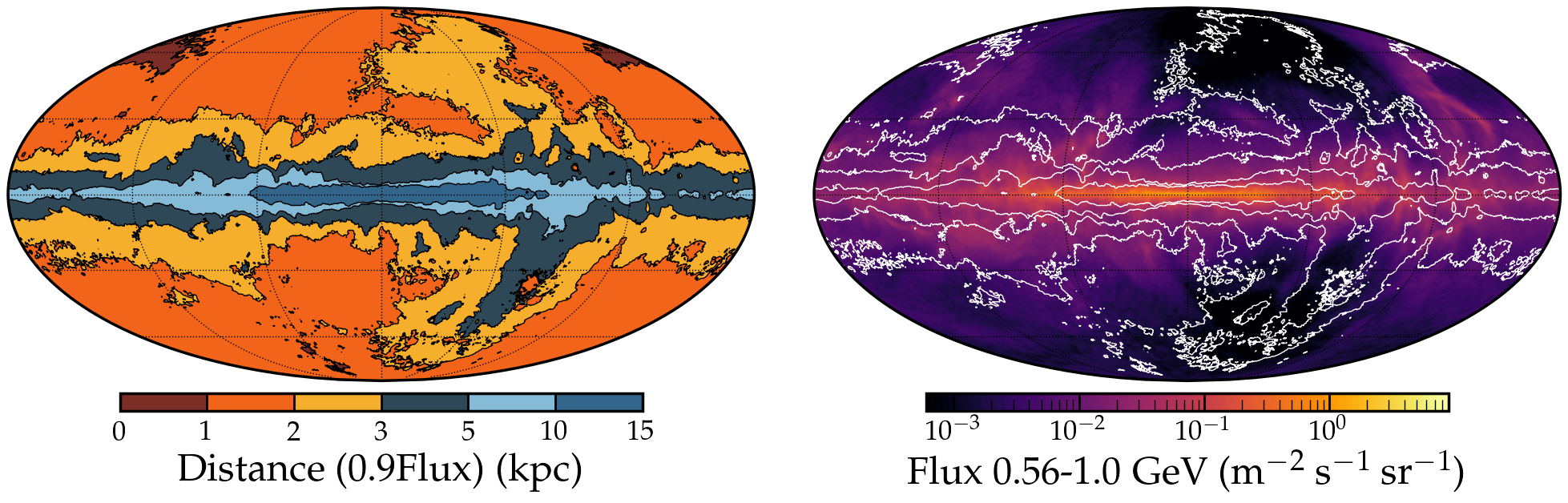}
  \caption{Same as Fig. \ref{fig:d90_flux}, but for different bubbles. The top right, bottom left, and bottom right panels correspond to the bubbles 1, 2, and 5. The top left bubble has coordinates $x=-5~\mathrm{kpc},y=6.4~\mathrm{kpc}$, but is not included in Fig. \ref{fig:d90_flux}. As before, we see that local emission dominates at higher galactic latitudes.}
  \label{fig:otherbubbles}
\end{figure*}

\begin{figure*}
    \centering
    \includegraphics[width=0.49\textwidth]{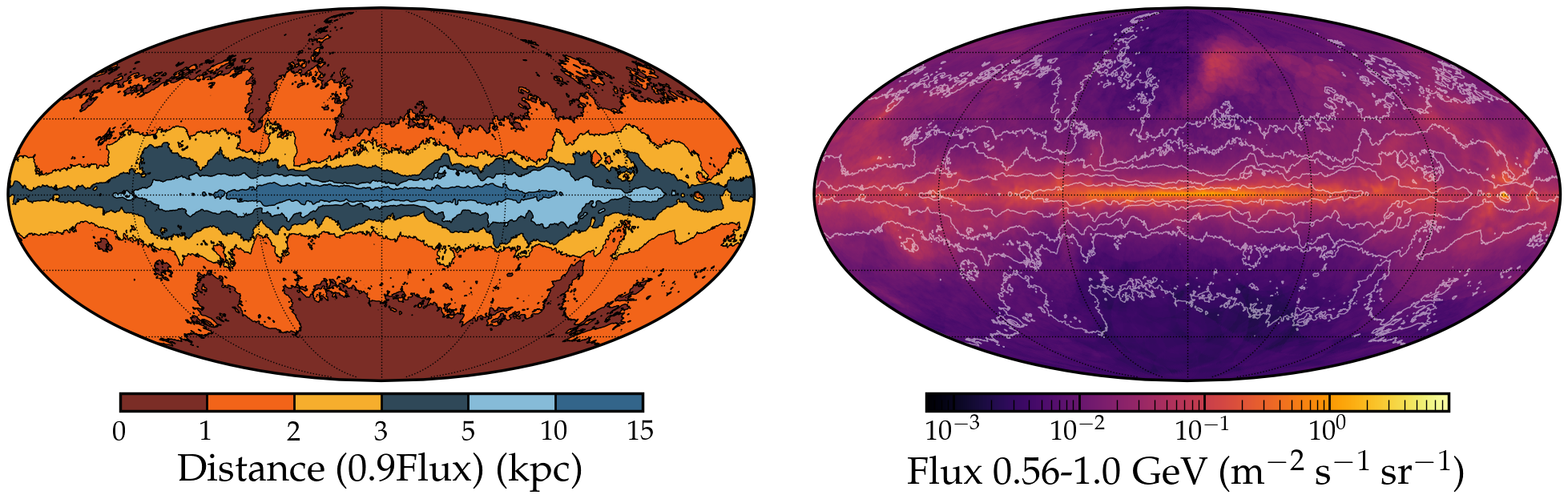}
    \hfill
    \includegraphics[width=0.49\textwidth]{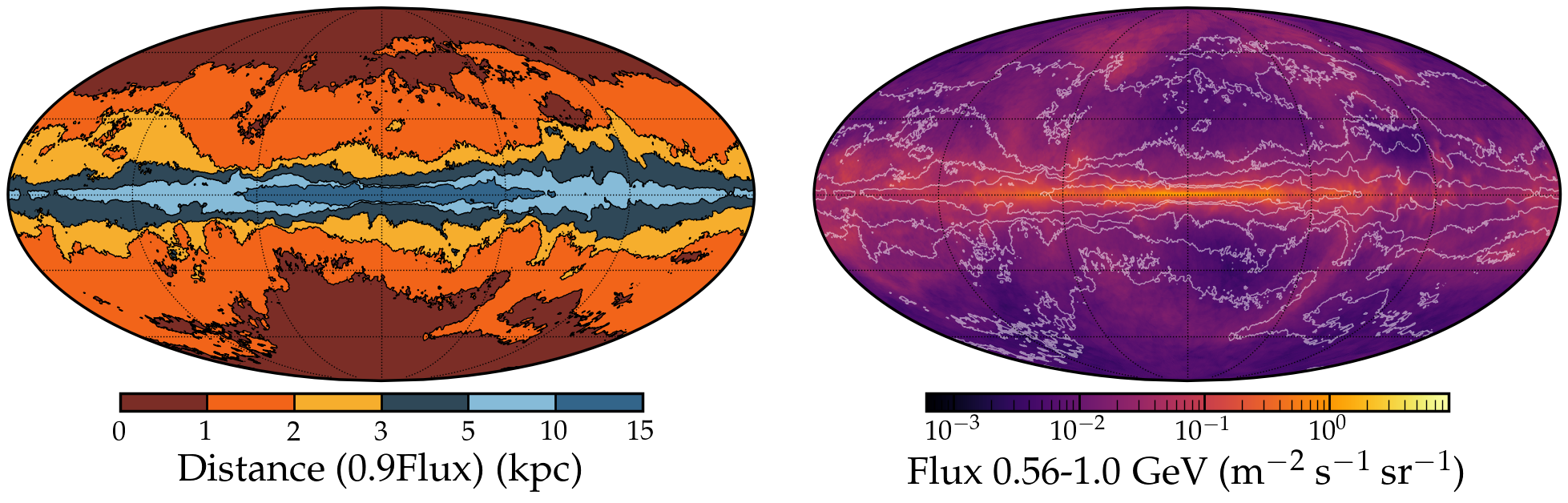} 
    \caption{Same as Fig. \ref{fig:d90_flux}, but for two random positions in the galaxy, that do not correspond to bubbles. }
    \label{fig:d90_nonbubbles}
\end{figure*}

In Fig. \ref{fig:otherbubbles} we show the same sky projections as in Fig.~\ref{fig:d90_flux} but for different observer positions. In all cases there is visible contribution from local emission, showing that our conclusions regarding the latitude-dependent integration-length are robust. In particular there are cases where the immediate surroundings (0--1 kpc from the observer) set the emission in the sky. Our conclusions also hold even in the case where an observer is placed in a nonbubble location, two examples of which are shown in Fig. \ref{fig:d90_nonbubbles}.

\section{CRMHD versus CRMHD-low}\label{app:crmhdlow}
\begin{figure*}[htbp]
    \centering
    \includegraphics[width=0.49\textwidth]{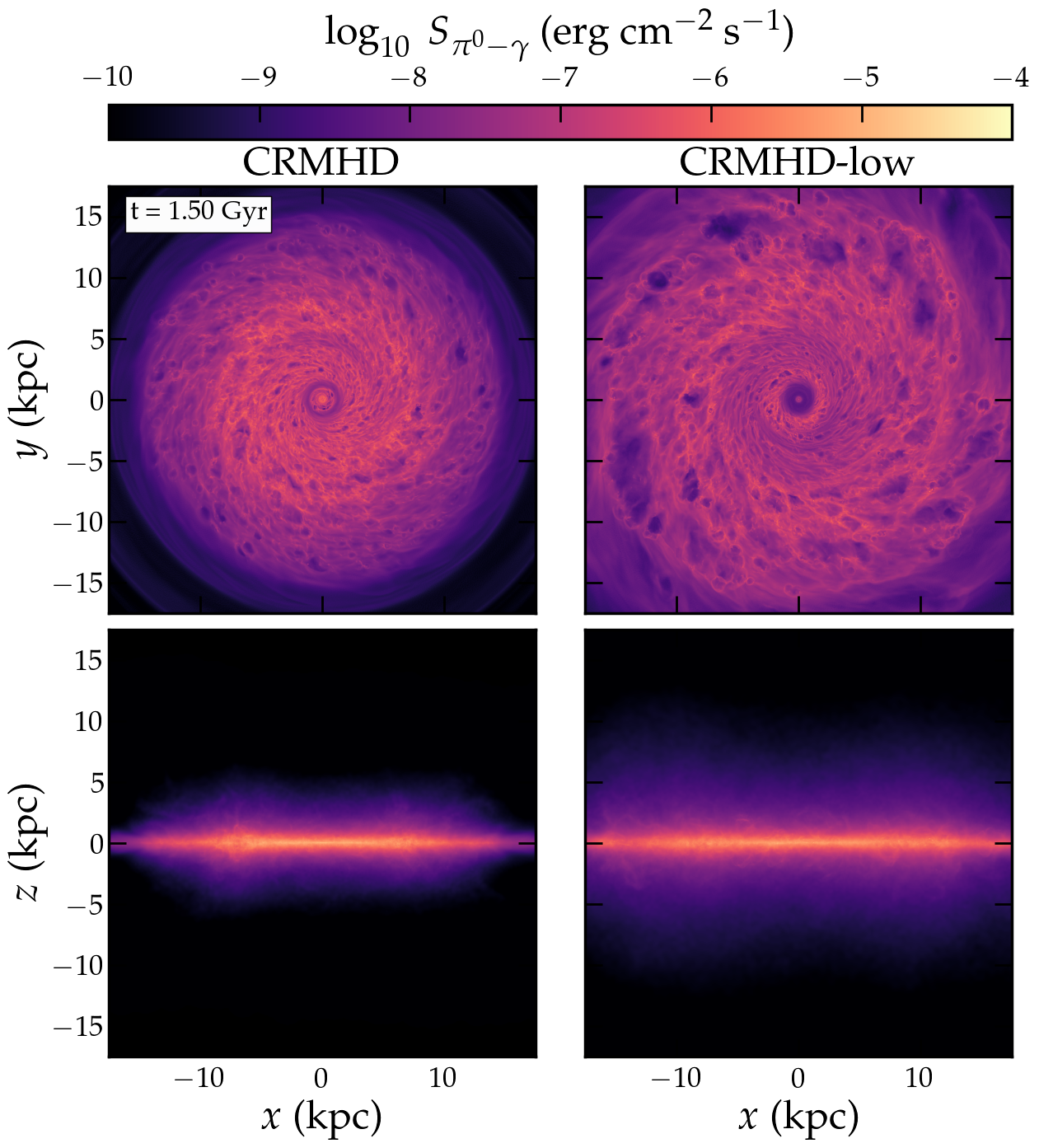}
    \hfill
    \includegraphics[width=0.49\textwidth]{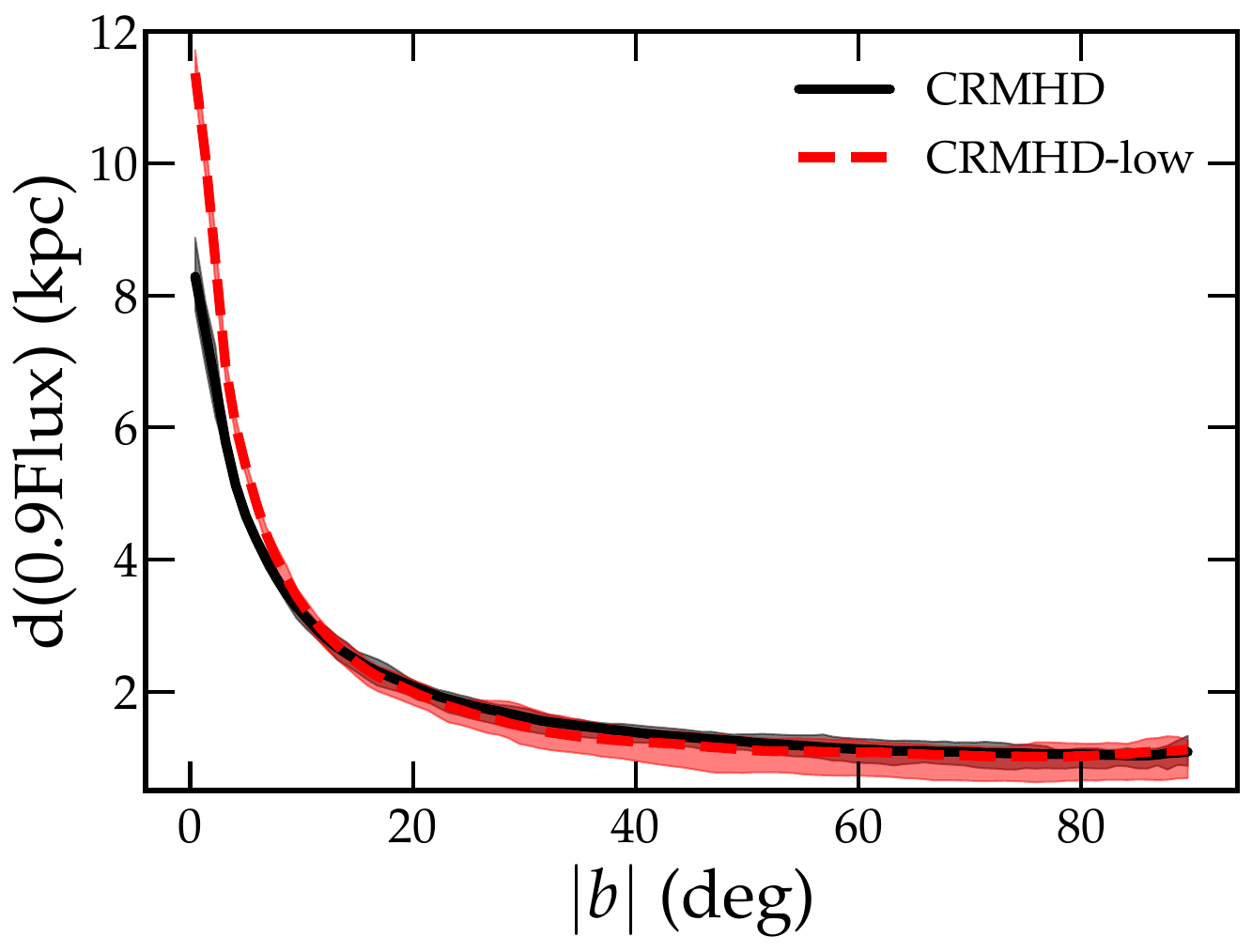}
    \caption{Left: Face-on and edge-on projections of the gamma-ray emissivity for our simulations CRMHD (left column) and CRMHD-low (right column). Right: Same as Fig. \ref{fig:d90_b}, but comparing CRMHD and CRMHD-low. Due to the initially weaker magnetic field, CRMHD-low has a larger disk, meaning that for small galactic latitudes most of the emission is coming from farther away than in CRMHD. At higher latitudes the two simulations behave the same. }
    \label{fig:appB}
\end{figure*}

For the analysis we chose to focus on one of our simulations, CRMHD. For the sake of completion and to show the robustness of our results we also show some results from our other simulation, CRMHD-low, which was initialized with 0.1\% of the magnetic field strength of CRMHD, but is in other regards identical. The main difference between the two simulations is that CRMHD-low has a larger star-forming disk as a result of the weaker magnetic support, as seen in the face-on panels on the left in Fig. \ref{fig:appB}. This is also reflected in the $d(0.9\mathrm{Flux})$-vs-$|b|$ plot in the right panel. Both simulations follow the same relationship, except at small galactic latitudes where the emission originates larger distances in CRMHD-low, due to the larger disk.

\end{appendix}

\end{document}